\documentclass[final,10pt,twocolumn]{elsarticle}
\usepackage{amssymb,pifont,latexsym}
\usepackage{amsmath,amsbsy,bbm}
\usepackage{epsfig,bm,color,bbold}
\usepackage{graphicx,url,hyperref}

\DeclareMathOperator{\sgn}{sign}

\journal{Annals of Physics}

\begin{document}

\def\a{{\alpha}}
\def\be{{\beta}}
\def\d{{\delta}}
\def\D{{\Delta}}
\def\p{{\pi}}
\def\e{{\varepsilon}}
\def\g{{\gamma}}
\def\k{{\kappa}}
\def\l{{\lambda}}
\def\L{{\Lambda}}
\def\m{{\mu}}
\def\o{{\omega}}
\def\S{{\Sigma}}

\def\ol#1{{\overline{#1}}}
\def\c#1{{\mathcal{#1}}}
\def\b#1{{\bm{#1}}}
\def\eqref#1{{(\ref{#1})}}

\begin{frontmatter}%

\title{Contact Interactions, Self-Adjoint Extensions, and Low-Energy Scattering}

\author[label1]{Daniel~R.~DeSena}
\ead{ddesena000@citymail.cuny.edu}
\author[label1,label2]{Brian~C.~Tiburzi\corref{cor1}}
\cortext[cor1]{Corresponding Author}
\ead{btiburzi@ccny.cuny.edu} 

\affiliation[label1]{organization={Department of Physics, The City College of New York},
            city={New York},
            state={NY},
            postcode={10031}, 
            country={USA}}

\affiliation[label2]{organization={Graduate School and University Center, The City University of New York},
            city={New York},
            state={NY},
            postcode={10016}, 
            country={USA}}

\begin{abstract}
Low-energy scattering 
is well described by the effective-range expansion. 
In quantum mechanics, 
a tower of contact interactions can generate terms in this expansion after renormalization. 
Scattering parameters are also encoded in the self-adjoint extension of the Hamiltonian. 
We briefly review this well-known result for two particles with 
$s$-wave interactions using impenetrable self-adjoint extensions,  
including the case of harmonically trapped two-particle states. 
By contrast, 
the one-dimensional scattering problem is surprisingly intricate.  
We show that the families of self-adjoint extensions correspond to a coupled system of symmetric and antisymmetric outgoing waves, 
which is diagonalized by an 
$SU(2)$
transformation
that accounts for mixing and a relative phase. 
This is corroborated by an effective theory computation that includes all four energy-independent contact interactions.
The equivalence of various one-dimensional contact interactions is discussed and scrutinized
from the perspective of renormalization. 
As an application, 
the spectrum of a general point interaction with a harmonic trap is solved in one dimension.
\end{abstract}

\begin{keyword}
Contact interactions 
\sep
Effective theories
\sep
Quantum mechanics
\sep
Renormalization
\sep
Robin boundary conditions
\sep
Scattering theory
\sep
Schr\"odinger equation 
\sep 
Self-adjoint extensions

\end{keyword}%
\end{frontmatter}%

\section{Overview}

Scattering of particles with short-range interactions
is well described at low energies by the effective-range 
expansion%
~\cite{Bethe:1949yr}.
From a modern perspective, 
this expansion emerges from a systematic treatment of 
contact interactions 
within the context of effective quantum field theories. 
For systems exhibiting resonant interactions near threshold, 
there has been a wealth of investigations and applications in atomic and nuclear physics%
~\cite{Bedaque:2002mn,Epelbaum:2008ga,RevModPhys.82.1225,Hammer:2019poc}.

Quantum mechanics with contact interactions has a long history, 
however, 
starting with the pseudo-potential method, 
which is rooted in self-adjoint extensions of Hamiltonian operators. 
For a discussion of the connection between these and effective field theory, 
see
Refs.~\cite{Jackiw:1991je,vanKolck:1998bw}. 
Further elaboration of the connection between self-adjoint extensions and one-dimensional contact 
interactions is the main focus of the present work. 
Early on, 
von Neumann recognized that physical observables are described by 
Hilbert-space operators that are not only Hermitian, 
but also self adjoint%
~\cite{vonNeumann}. 
While most discussion has been centered in the mathematical physics community%
~\cite{Reed:1975uy}, 
differential operators with point interactions are routinely used to obtain 
exactly soluble quantum mechanical models%
~\cite{Albeverio}.
One-dimensional point interactions have been discussed from a variety of perspectives%
~\cite{Seba:1986,Carreau:1990wh,Carreau:1991us,CHERNOFF199397,Coutinho:1997ab,FABCoutinho_1999}, 
with an increasing slant toward physics phenomenology. 
More recent work has investigated novel applications concerning confined systems 
\cite{Al-Hashimi:2011xuv,Al-Hashimi:2012pod}. 
A particularly lucid introduction to self-adjoint extensions of Hamiltonians is contained in
Ref.~\cite{Al-Hashimi:2013rra}, 
which additionally details an interesting application to supersymmetric 
quantum descendants.

Contact operators afford a complementary description of short-range interactions. 
In one dimension, 
the 
$\d(x)$
potential is a staple problem of introductory quantum mechanics.  
Higher-dimensional contact operators, 
however, 
require regularization and renormalization.%
\footnote{
For an introduction to regularization and renormalization in the context of 
quantum mechanics, 
see Ref.~\cite{Lepage:1997cs}.
} 
The very singular
$\d''(x)$
potential, 
for example,
has been considered using various techniques.
In Ref.~\cite{Al-Hashimi:2016apw}, 
for example,
the interaction was treated by means of a hard momentum cutoff,%
\footnote{
A different treatment of the 
$\d''(x)$
interaction based on the theory of discontinuous distributions
is given in
Ref.~\cite{JAROSZ2021168617}.
These results can be obtained with the 
NDR scheme described in%
~\ref{s:DimReg}.
} 
for which the equivalence to a delta-function potential has
been exhibited in the strict limit of an infinite cutoff. 
This quantum mechanical result demonstrates a well-known feature of renormalization theory:
the 
$\d''(x)$
potential is an irrelevant operator, 
and the lower dimensional 
$\d(x)$
potential is a relevant operator. 
No symmetry protects the generation of the delta-function potential
under renormalization group evolution.%
\footnote{
Symmetry properties of second-derivative contact interactions are discussed in 
Sec.~\ref{s:C1D}. 
In particular, 
Eq.~\eqref{eq:relation}
shows that the renormalization of 
$\d''(x)$
is complicated by its coupling to both symmetric and antisymmetric waves. 
Furthermore, 
only the coupling to antisymmetric waves is described by a self-adjoint operator. 
}

A more curious interaction is provided by the 
$\d'(x)$
potential,%
\footnote{
When discussing point interactions, 
one must be careful to distinguish this interaction 
$\d'(x) = \frac{d}{dx}\d(x)$, 
from the unfortunately named 
$\d'$-interaction. 
For the latter, 
the prime is used in the sense of an alternative rather than to denote differentiation.
} 
which is a marginal operator. 
In one dimension, 
the Hamiltonian with such a potential has a classical scale symmetry, 
for which the generation of a
$\d(x)$
operator under renormalization group evolution would ordinarily be forbidden. 
There is, 
however, 
a scale anomaly in relativistic quantum field theory%
~\cite{COLEMAN1971552,Jackiw:1972cb}, 
which also appears in non-relativistic quantum mechanics, 
see, 
for example, 
Refs.~\cite{Bergman:1991hf,Camblong:2000qn}.
The classical symmetry is not a symmetry at the quantum level, 
and the relevant 
$\d(x)$
operator is generated by renormalization group evolution. 
This is a rephrasing of the results originally obtained in 
Ref.~\cite{SEBA1986}. 
The result is also illustrated as a preliminary example in 
Ref.~\cite{Al-Hashimi:2015nva}, 
and given further exposition using the language of effective field theory in 
Ref.~\cite{Camblong:2019bfr}.

From a renormalization group perspective, 
it does not make sense to consider theories with marginal and irrelevant operators
without also including the relevant 
$\d(x)$
potential. 
Previous work demonstrates that it will be generated under renormalization group evolution. 
As we show below, 
its inclusion is additionally justified on the grounds of renormalizability;
otherwise, 
the scattering matrix is renormalization scale dependent, 
albeit possessing a finite limit as the renormalization scale is taken to infinity.   
This treatment, 
moreover, 
accommodates results obtained from defining singular interactions from discontinuous distributions%
~\cite{KURASOV1996297}, 
such as those of 
Refs.~\cite{GADELLA20091310,Gadella2011}, 
which are particular examples of the most general self-adjoint extension of the free-particle Hamiltonian 
on the punctured line.

Our presentation is organized as follows. 
First in 
Sec.~\ref{sec:swave}, 
we review the impenetrable self-adjoint extensions that describe the relative radial problem of 
two particles with a short-range $s$-wave interaction.  
Robin boundary conditions are shown to incorporate the scattering length. 
This well-known result is then applied to reproduce the spectrum of harmonically trapped two-particle states.  
In Sec.~\ref{sec:point}, 
self-adjoint extensions on the punctured line are reviewed. 
These give rise to the most general one-dimensional point interaction, 
for which we detail its properties under symmetry transformations.  
Scattering from a point interaction is solved in terms of the 
$S$-matrix, 
with the general point interaction exhibiting partial-wave mixing and a relative phase.  
The analogues of partial waves in one dimension are symmetric and antisymmetric waves;
scattering of such parity waves is detailed in 
\ref{s:parity}. 
Various limiting cases of the scattering solution for a point interaction are discussed, 
including a few curious results.

The task of reproducing the behavior of the general point interaction from an effective theory is undertaken in 
Sec.~\ref{s:C1D}. 
All contact interactions up to second-derivative order are enumerated, 
along with their properties under symmetry transformations.  
The scattering problem is solved in momentum space, 
for which technical details concerning regularization appear in 
\ref{s:DimReg}. 
Renormalized results for parity-even and parity-odd interactions are contrasted in two regularization schemes.
Using the four energy-independent contact interactions with na\"ive dimensional regularization, 
we obtain the relations between coefficients of contact operators and the self-adjoint extension 
parameters.  
A final application is pursued in Sec.~\ref{s:SHO1D}, 
where the one-dimensional problem of two harmonically trapped particles with a general point interaction is solved. 
Our results indirectly confirm the quantum scale anomaly. 
A final summary of key results and remaining questions is given in 
Sec.~\ref{s:summy}.

\section{Impenetrable Self-Adjoint Extensions}
\label{sec:swave}

We use the relative problem of two particles with $s$-wave interactions to review the 
impenetrable self-adjoint extension. 
At sufficiently low energies, 
such short-range interactions can be parameterized by a tower of contact interactions 
\begin{equation}
V_S(r)
=
c_0 \, \delta^{(3)}(\vec{r}\,)
+ 
c_2 \left\{  \nabla^2 , \delta^{(3)} (\vec{r}\,) \right\}
+ 
\cdots
\label{eq:contact}
,\end{equation}
regardless of the system under consideration. 
This is because the wavelength of the probe cannot resolve the detailed short-range structure of the interaction. 
In Eq.~\eqref{eq:contact}, 
we have exhibited the leading energy-independent contact interaction, 
along with the first energy-dependent term. 
In coordinate space, 
this is a derivative expansion
with omitted terms 
having at least four derivatives. 
Due to ultraviolet divergences, 
these interactions require 
regularization, 
and the parameters
$c_0$, 
$c_2$, 
$\cdots$,
consequently become scale- and scheme-dependent 
running couplings. 
Physical properties 
are rendered scale and scheme independent after enforcing renormalization conditions. 
The short-range interactions are assumed strong,
so that 
$V_S(r)$
is not amenable to perturbation theory.

In a low-energy description, 
exclusion of the origin removes the short-range interaction.  
With 
$r \neq 0$, 
the reduced radial Hamiltonian for 
$s$-waves is
\begin{equation}
H_0 \big(r{>}0 \big)
=
- \frac{1}{2m} \frac{d^2}{dr^2} + V_L(r)
,\end{equation}
where 
$m$
is the reduced mass, 
$\hbar = 1$
in the units we employ throughout, 
and we have additionally included a rotationally invariant long-range potential $V_L(r)$.%
\footnote{
This form of the Hamiltonian 
assumes that the separation of long- and short-range contributions emerges in the low-energy limit, 
and is discussed further below. 
} 
Extension of the Hamiltonian to 
$r =0$
can be achieved by general principles, 
namely by requiring that 
$H_0$
be self adjoint. 
On the half line
$r>0$, 
the physical requirement resulting from the self-adjoint extension is that the probability current vanishes at the origin.
This is the so-called impenetrable self-adjoint extension. 
The reduced-radial energy eigenfunctions 
$u_k(r)$, 
satisfy the eigenvalue equation
\begin{equation}
H_0 \big(r{>}0\big) \, u_k(r) = E \, u_k(r)
\label{eq:EXCL}
,\end{equation} 
with the energy eigenvalue written as 
$E = \frac{k^2}{2m}$, 
where
$k>0$
is the scattering momentum.
For such solutions,
the radial probability current is 
\begin{equation}
J_k(r) = \frac{1}{m} \mathfrak{Im} \big[ u^*_k(r) \, u'_k(r) \big]
,\end{equation}
where the prime denotes differentiation with respect to 
$r$. 
The most general boundary condition leading to a vanishing probability current
\begin{equation}
J_k(\e) = 0
\quad
\text{for}
\quad 
\e \to 0^+
,\end{equation}
is the homogeneous Robin boundary condition
\begin{equation}
u'_k(\e) - \be \, u_k(\e) = 0
\label{eq:Robin}
,\end{equation}
where 
$\be$
is the real-valued self-adjoint extension parameter. 
As the boundary condition must be energy independent to ensure orthogonality of the eigenstates,%
\footnote{
Related problems with Hermiticity were noted long ago in the context of
energy-dependent pseudo-potentials%
~\cite{PhysRev.105.767}.
For an analysis of energy-dependent point interactions, 
see 
Refs.~\cite{Coutinho,doi:10.1139/p06-086}.
Because such interactions are not self-adjoint, 
they lie outside the scope of our investigation. 
} 
the self-adjoint extension accounts for the energy-independent 
contact interaction in 
Eq.~\eqref{eq:contact}.

\subsection{Scattering}        %

For the case of scattering off a short-range potential, 
we assume the
long-range potential vanishes.
On account of unitarity, 
the $s$-wave scattering amplitude takes the general form
\begin{equation}
f_0(k)
=
\frac{1}{k \cot \delta_0 - i k}
\label{eq:kcotd}
,\end{equation}
where 
$\d_0 = \d_0(k)$
is the 
$s$-wave phase shift. 
At low energies, 
one has a well-behaved expansion of the particular combination
\begin{equation}
k \cot \delta_0 
=
 - \frac{1}{a} + \frac{1}{2} r_0 \, k^2 + \c O(k^4)
\label{eq:ERE}
,\end{equation}
where
$a$ 
is the scattering length 
(using the nuclear physics sign convention)
and 
$r_0$
is the effective range. 
Higher-order terms become relevant as the energy increases.

To describe $s$-wave scattering at low energies, 
we exclude the origin, 
for which the reduced radial Hamiltonian is simply
\begin{equation}
H_0\big(r{>}0\big) = - \frac{1}{2m} \frac{d^2}{dr^2}
\label{eq:FREE}
.\end{equation} 
Including short-range interactions can be achieved with a 
self-adjoint extension of the kinetic energy operator on the half line. 
With the standard elastic scattering solution written in terms of the $s$-wave phase shift 
\begin{equation}
u_k(r) = N \, \sin( k r + \delta_0)
,\end{equation}
one obtains the self-adjoint extension of 
$H_0$
by enforcing the Robin boundary condition in Eq.~\eqref{eq:Robin}. 
This immediately leads to
\begin{equation}
\be 
= 
- \frac{1}{a}
\label{eq:betaa}
.\end{equation}
The scattering-length contribution to the effective-range expansion
Eq.~\eqref{eq:ERE}
emerges from the self-adjoint extension of the reduced radial Hamiltonian.

With divergences stemming from 
$r=0$
excluded, 
the self-adjoint extension of the Hamiltonian sidesteps regulating the short-range interaction.
Including the origin, 
on the other hand,
generally leads to power-law divergences from contact interactions. 
Once they are regulated, 
renormalization is carried out by matching the low-energy behavior of the scattering 
amplitude
Eq.~\eqref{eq:kcotd}.
In effective field theory,
power-counting schemes have been devised to carry out such matching in systematically improvable ways. 
The above toy model with a large scattering length, 
for example, 
was considered in  
Refs.~\cite{Kaplan:1998tg,Kaplan:1998we} 
as a prelude to addressing the two-nucleon system.
Large range corrections can additionally be summed, 
and higher-order calculations 
simplified by employing an efficacious basis for higher-dimensional contact interactions%
~\cite{Beane:2000fi}.
Such energy-dependent corrections, 
however, 
lie outside our consideration of the self-adjoint extension of 
Eq.~\eqref{eq:FREE}.

\subsection{Harmonic Confinement}                         %
\label{s:SHO3}                                                           %

\begin{figure}[tb]
\begin{center}
\includegraphics[scale=0.725]{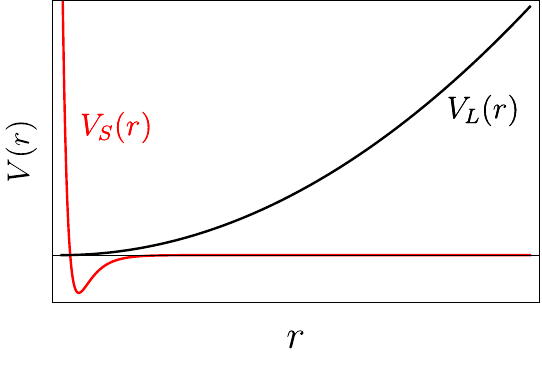}
\
\includegraphics[scale=0.725]{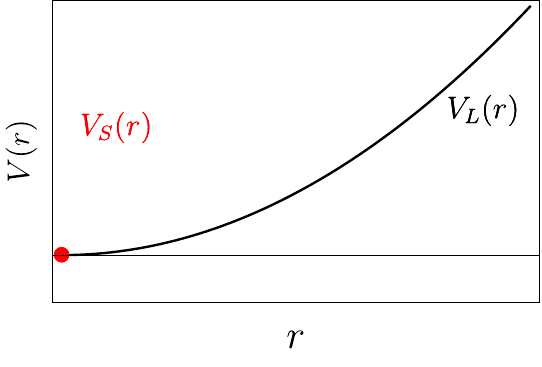}
\caption{Schematic atom-atom radial potential in terms of long- and short-range contributions.
On top is a depiction of a short-range potential and a long-range harmonic trap. 
The harmonic potential approximately vanishes at short range.
On the bottom, the short-range potential has been replaced by a contact interaction
assuming the separation of length scales.
}
\label{f:SHO}
\end{center}
\end{figure}

In an isotropic harmonic trap, 
the long-range potential is
$V_L(r) = \frac{1}{2} m \o^2 r^2$. 
With the origin excluded, 
the reduced radial Hamiltonian for $s$-waves reads
\begin{equation}
H_0\big(r{>}0\big)
=
- \frac{1}{2m} \frac{d^2}{dr^2} + \frac{1}{2} m \o^2 r^2
\label{eq:HSHO}
.\end{equation}
In parallel to the previous case, 
we obtain the solution for the energy eigenfunction 
$u_E(r)$
on the half line
$r > 0$,
and then enforce the Robin boundary condition 
Eq.~\eqref{eq:Robin}
to obtain the self-adjoint extension of 
$H_0$.
There is a crucial feature relevant to this analysis. 
The harmonic potential vanishes at the origin;
thus,
it is completely absent from the low-energy description of the short-range potential. 
Consequently, 
the self-adjoint extension parameter
$\be$
is independent of the harmonic frequency
$\o$, 
and is determined from 
Eq.~\eqref{eq:betaa}. 
The problem of harmonically trapped particles with 
$s$-wave interactions at low energies should have a clear separation between the long-range and short-range effects,%
\footnote{
This assumption can be violated by increasing the strength of the confining potential, 
such as was observed from detailed microscopic calculations of alkali atoms in a strong trap%
~\cite{PhysRevA.61.063416}.
}
as shown in 
Fig.~\ref{f:SHO}.
 
In the region 
$r > 0$, 
the normalizable solution to the radial equation posed by 
$H_0$
in 
Eq.~\eqref{eq:HSHO} 
is given by
\begin{equation}
u_E(r) = N \ U \hskip-0.25em \left(-\tfrac{E}{\o},\sqrt{2m\o} \, r \right)
,\end{equation}
where 
$U(a,z)$
is a parabolic cylinder function, 
which accordingly satisfies
$U \big(a,z \, {\to}\, \infty \big) = 0$.
Enforcing the Robin boundary condition at the origin leads to 
\begin{equation}
\be = \sqrt{2 m \o} \ \frac{U'\big({-}\tfrac{E}{\o},0\big)}{U\big({-}\tfrac{E}{\o},0\big)}
,\end{equation}
where the prime denotes differentiation with respect to the second argument.  
Values of the parabolic cylinder function and its derivative at the origin 
are%
~\cite[Eqs.~(12.2.6) and (12.2.7)]{NIST:DLMF}
\begin{equation}
U(a,0)
=
\frac{2^{-\frac{1}{4}-\frac{a}{2}} \sqrt{\p}}{\Gamma\left(\tfrac{3}{4} + \tfrac{a}{2}\right)}
\, \, \text{ and } \, \,
U'(a,0)
=
- \frac{2^{\frac{1}{4} - \frac{a}{2}} \sqrt{\p}}{\Gamma\left( \tfrac{1}{4} + \frac{a}{2} \right)}
\label{eq:UU'}
.\end{equation}
Combining these results, 
we arrive at the spectrum condition 
\begin{equation}
- \frac{1}{a}
= 
 - 2 \sqrt{m \o} \
\frac{\Gamma\left(\tfrac{3}{4} - \tfrac{E}{2 \o}\right)}{\Gamma\left( \tfrac{1}{4} - \frac{E}{2\o} \right)}
\label{eq:TRAP}
.\end{equation}
Without explicitly regulating divergences,
the self-adjoint extension of the Hamiltonian introduces the relevant physics, 
which is the
$s$-wave scattering length in the absence of the harmonic trap. 
The crucial feature is the vanishing of the oscillator potential near the origin, 
as it leads to 
separation between the long-range and short-range physics
in the low-energy limit.

The above transcendental equation for the spectrum was first derived in 
Ref.~\cite{Busch:1998} 
using a pseudo-potential method. 
That approximation was later scrutinized by a detailed microscopic calculation of the short-range interaction
in a strong trap%
~\cite{PhysRevA.61.063416}. 
Various investigations%
~\cite{PhysRevA.65.043613,PhysRevA.65.052102,PhysRevA.66.013403}
improved upon the approximation using an effective scattering-length model 
or an energy-dependent pseudo-potential. 
These approaches lead to the replacement of
$- \frac{1}{a}$
in 
Eq.~\eqref{eq:TRAP}
with the quantity
$k \cot \d_0$. 
An effective field theory calculation of this result appears to have been carried out first in 
Ref.~\cite{Mehen:2007dn}.
As the energy dependence of 
$k \cot \d_0$
requires energy-dependent point interactions,
the improved formula lies outside our consideration of the self-adjoint extension of 
Eq.~\eqref{eq:HSHO}.

\section{General Point Interaction in One Dimension}
\label{sec:point}

The above examples are one-dimensional problems formulated on the half line 
$r > 0$, 
and utilize the impenetrable self-adjoint extensions to 
$r = 0$.
By contrast, 
one-dimensional problems on the punctured line
$x \in \mathbb{R}/\{0\}$
actually represent a greater challenge, 
for which we begin our investigation with a vanishing long-range potential. 
The self-adjoint extensions now allow for transmission through the origin in addition to reflection. 
These extensions we discuss in 
Sec.~\ref{s:SAE1D}, 
and the scattering problem is solved in 
Sec.~\ref{s:SAEscattering}. 
Finally, 
the solution is investigated in various limiting cases in 
Sec.~\ref{s:SAElimits}.

\subsection{Self-Adjoint Extension for a Point Interaction in One Dimension}
\label{s:SAE1D}

For the one-dimensional kinetic energy operator
\begin{equation}
H = - \frac{1}{2m} \frac{d^2}{dx^2}
,\end{equation}
defined on the punctured line
$x \in \mathbb{R}/\{0\}$, 
the most general joining conditions for the coordinate-space wavefunction 
$\psi(x)$
across the origin 
$\e \to 0^+$
are written as
\begin{equation}
\begin{pmatrix}
\psi'(\e)
\\
\psi(\e)
\end{pmatrix}
=
\c M
\begin{pmatrix}
\psi'(-\e)
\\
\psi(-\e)
\end{pmatrix}
\label{eq:SAE1D}
,\end{equation}
where, 
in the notation of 
Ref.~\cite{Coutinho:1997ab},
the transfer matrix 
$\c M$
takes the form
\begin{equation}
\c M 
= 
e^{i \phi} 
\begin{pmatrix} 
\a & \be \\ \d & \g 
\end{pmatrix}
\label{eq:M}
.\end{equation}
In the parameterization of 
$\c M$, 
all parameters are real valued 
and satisfy the constraint
\begin{equation}
\a \, \g - \be \, \d = 1
\label{eq:det}
.\end{equation} 
To avoid redundancy, 
the phase is restricted to
$\phi \in ( - \frac{\p}{2}, \frac{\p}{2} \, ]$. 
The parameters 
$\a$
and
$\g$
are dimensionless, 
whereas 
$\be^{-1}$ 
and 
$\d$ 
are lengths. 
These joining conditions on the wavefunction and its derivative are a consequence of the self-adjoint extension of 
$H$. 
Physically, 
they enforce that the probability current is the same on both sides of the origin, 
at which the wavefunction need be neither continuous nor differentiable. 
This behavior at the origin is said to result from a point interaction. 

Properties of the general point interaction under parity, time-reversal, and scaling transformations 
were investigated in 
Ref.~\cite{Albeverio1998}. 
To describe the properties of 
$\c M$
under transformations, 
we find it convenient to employ the matrices
\begin{equation}
\S_1 
{=} 
\begin{pmatrix}
0 & 1 \\ 1 & 0 
\end{pmatrix}, 
\,\,\,
\S_2 
{=} 
\begin{pmatrix} 0 & - i \\ i & \phantom{-} 0 \end{pmatrix},
\,\,\,
\S_3
{=} 
\begin{pmatrix} 1 & \phantom{-} 0 \\ 0 & - 1\end{pmatrix}
\label{eq:Smatrices}
.\end{equation}
The joining conditions in 
Eq.~\eqref{eq:SAE1D}
can equivalently be written by traversing the origin in the opposite direction, 
for which we have
\begin{equation}
\begin{pmatrix}
\psi'(-\e)
\\
\psi(-\e)
\end{pmatrix}
=
\c M^{-1}
\begin{pmatrix}
\psi'(\e)
\\
\psi(\e)
\end{pmatrix}
\label{eq:SAE1Dreflected}
.\end{equation}
The inverse matrix 
$\c M^{-1}$
satisfies the conjugacy relation
\begin{equation}
\c M^{-1}
=
\Sigma_2 \, \c M^\dagger \, \Sigma_2
\label{eq:Minv}
.\end{equation}
Consequently, 
the matrix 
$\c M$
will be be unitary provided
$[ \c M, \S_2 ] =0$. 
For arbitrary 
$\phi$,%
\footnote{
There is another possibility exclusive to the particular value
$\phi = \frac{\p}{2}$. 
One must first redefine the matrix 
$\c M$
to be dimensionless, 
for example, 
by multiplying 
$\psi'(\pm\e)$
by a characteristic length in Eq.~\eqref{eq:SAE1D}. 
Unitarity of   
$\c M$
is then possible for any value of 
$\xi$, 
where
$\cos \xi \equiv \a = \g$
and
$\sin \xi \equiv \d = - \be$.  
The latter condition is only possible when 
$\be$ 
and 
$\d$ 
are dimensionless.
The value 
$\phi = \frac{\p}{2}$
is peculiar, 
as is briefly addressed in 
Sec.~\ref{s:TREI}. 
}
this can only be achieved by requiring 
$\be = \d = 0$
and
$\a = \g = \pm 1$,  
for which the matrix
$\c M$
is simply the identity matrix times a phase. 
In this special case, 
the wavefunction and its derivative are continuous up to (the same) phase, 
leading to a continuous logarithmic derivative at the origin.%

Under time reversal,
the wavefunction has the antilinear transformation
$\psi(x) \to \psi^*(x)$;
consequently, 
from 
Eq.~\eqref{eq:SAE1D}, 
the matrix 
$\c M$
also has an antilinear transformation 
$\c M \to \c M^*$. 
This transformation can be achieved simply by the replacement 
$\phi \to - \phi$, 
from which we infer that
$\phi$
is time-reversal odd 
and all other parameters are time-reversal even.

A parity reflection about the origin produces the interchange 
\begin{equation}
\begin{pmatrix}
\psi'(-\e)  
\\
\psi(-\e) 
\end{pmatrix}
\to 
\S_1 \S_3 \S_1
\begin{pmatrix}
\psi'(\e)  
\\
\psi(\e) 
\end{pmatrix}
,\end{equation}
which takes into account the behavior of the derivative
$\frac{d}{dx} \to - \frac{d}{dx}$
under spatial reflection. 
A parity transformation consequently affects the joining conditions through
\begin{equation}
\c M
\to
\S_3 \, \c M^{-1} \, \S_3 
.\end{equation}
In terms of parameters of the self-adjoint extension, 
parity invokes the transformation 
$\a \leftrightarrow \g$
and
$\phi \to - \phi$. 
Demanding that the joining conditions respect parity invariance leads to the conjugacy relation
\begin{equation}
\c M = \S_1 \, \c M^\dagger \, \S_1
,\end{equation}
where we have combined the parity transformation with the form of the inverse written in 
Eq.~\eqref{eq:Minv}.  
Not surprisingly, 
parity invariance of the point interaction requires 
$\a = \g$
and
$\phi = 0$. 
For a point interaction that is invariant after the combined parity and time-reversal 
($PT$) 
transformations, 
one requires 
$\c M^* = \S_1 \, \c M^\dagger \, \S_1$. 
This less restrictive condition is met for 
$\a = \g$,
but for arbitrary values of 
$\phi$.
The parameter 
$\phi$
is both parity and time-reversal odd, 
hence, 
$PT$
even.

Finally, 
under the scale transformation
$x \to \lambda \, x$, 
the wavefunction obeys
$\psi(x) \to \psi(\lambda x) =  \lambda^{1/2} \, \psi(x)$. 
The joining conditions consequently have the scale transformation 
\begin{equation}
\c M 
\to 
\begin{pmatrix}
1 & 0 \\
0 & \lambda^{-1}
\end{pmatrix}
\c M
\begin{pmatrix}
1 & 0 \\
0 & \lambda 
\end{pmatrix}
.\end{equation}
A scale-invariant point interaction thus requires 
$\be = \d = 0$, 
which is not surprising given that each of these parameters carries physical dimensions. 
In this case, 
the constraint in 
Eq.~\eqref{eq:det} 
requires 
$\g = \a^{-1}$,
for which 
$\a$
and
$\phi$
are free parameters. 
If one further demands parity invariance, 
then 
$\phi =0$
and
$\a = \pm 1$; 
or, 
if one demands
$PT$
invariance, 
then 
$\a = \pm 1$
for arbitrary 
$\phi$. 
Both of these cases correspond to a unitary matrix
$\c M$.

\subsection{Scattering From a Point Interaction}
\label{s:SAEscattering}

Having spelled out the possible self-adjoint extensions on the punctured line, 
it is elementary to solve the quantum mechanical scattering problem 
subject to 
Eq.~\eqref{eq:SAE1D}. 
Beyond amplitudes for reflection and transmission, 
we obtain the 
$S$-matrix in the partial-wave basis.
Details concerning scattering theory in one dimension are presented in%
~\ref{s:parity}. 
Note that for a point interaction, 
exclusion of the origin automatically puts one in the asymptotic region, 
where the solutions are free-particle waves.

For incoming right- and left-traveling waves
(denoted by 
$\pm$ superscripts), 
the reflected and transmitted amplitudes are%
~\cite{CHERNOFF199397,Coutinho:1997ab,FABCoutinho_1999}
\begin{eqnarray}
R^{(\pm)} 
&=& 
\frac{k^2 \d \pm i k (\a - \g) + \be}{k^2 \d  + i k (\a + \g) - \be}
,\notag \\
T^{(\pm)} 
&=& 
\frac{2 i k \ e^{\pm i \phi}}{k^2 \d  + i k (\a + \g) - \be} 
\label{eq:RT}
,\end{eqnarray}
respectively.
Note that these amplitudes are related by the parity transformation of the 
self-adjoint extension parameters from above, 
so that
$R^{(+)} \overset{P}{\to} R^{(-)}$
and
$T^{(+)} \overset{P}{\to} T^{(-)}$. 
Despite the lack of continuity and differentiability at the origin, 
probability is conserved
$|R^{(\pm)}|^2 + |T^{(\pm)}|^2 = 1$. 
This is a necessary consequence of the physics underlying the self-adjoint extension.

To simplify the resulting expressions below, 
note that all amplitudes in 
Eqs.~\eqref{eq:RT}
share the same denominator
\begin{equation}
\c D
\equiv
k^2 \d  +  i k (\a + \g) - \be 
=
\d \, (k - i \k_+ ) (k - i \k_-)
\label{eq:Den}
,\end{equation}
which has (imaginary) roots
\begin{equation}
\k_\pm
=
\frac{-(\a+\g) \pm \sqrt{(\a - \g)^2 + 4\,}}{2 \d}
\label{eq:roots}
.\end{equation}
Using the reflected and transmitted amplitudes, 
the partial-wave 
$S$-matrix 
Eq.~\eqref{eq:SS}
for the general point interaction can be determined. 
It can be cast in the form
\begin{equation}
\mathbb{S} = \ol T \ \mathbb{1} + \frac{\vec{\c B} \cdot \vec{\S}}{\c D}
,\end{equation}
where we have repurposed the matrices 
$\vec{\S} = (\S_1, \S_2, \S_3)$
defined in 
Eq.~\eqref{eq:Smatrices}, 
$\mathbb{1}$
denotes the identity matrix, 
and the vector
\begin{equation}
\vec{\c B} = \Big( - 2 k \sin \phi, \, -  k (\a - \g), \, \be + k^2 \d \Big)
\label{eq:Bvec}
,\end{equation}
has three real components. 
The mixing of partial waves due to a parity and time-reversal breaking point interaction is therefore
mathematically the problem of spin-half in a magnetic field. 
A non-vanishing component of the field along the 
second direction leads to breaking of parity symmetry, 
while a component along the 
first direction leads to breaking of both parity and time-reversal.

Writing the mathematical analogue of the magnetic field in terms of its magnitude and direction
$\vec{\c B} = | \vec{\c B} \, | 
\, \hat{\c B}$, 
the eigenvalues of the 
$S$-matrix are thus 
$\ol T \pm | \vec{\c B} \, | / \c D$. 
Using the roots defined in 
Eq.~\eqref{eq:roots}, 
we can express the magnitude squared as
\begin{equation}
| \vec{\c B} \, |^2 
= 
\d^2 (k^2 + \k_+^2)(k^2+\k_-^2) - 4 k^2 \cos^2 \phi
,\end{equation}
from which the
$S$-matrix
eigenvalues can be written in the form
\begin{equation}
e^{2 i \d_\pm}
=
\frac{2 i k \cos \phi \pm | \vec{\c B} \, |}
{\d \, (k - i \k_+ ) (k - i \k_-)}
\label{eq:EVPhases}
.\end{equation}
Explicit computation confirms that these eigenvalues are unimodular, 
in accordance with the unitarity of 
$\mathbb{S}$. 
For this reason, 
the eigenvalues have been written in terms of phases angles
$\d_\pm$.

The direction of 
$\vec{\c B}$
can be expressed as the location of a point on a unit sphere. 
Instead of the azimuthal angle 
$\varphi$, 
it is convenient to use a rotated version
$\Phi = \varphi + \frac{\p}{2}$
and the express the direction as
\begin{equation}
\hat{\c B}
= 
\big(
\sin \Phi \sin\Theta, \, - \cos \Phi \sin \Theta,  \, \cos \Theta \big)
\label{eq:Bhat}
.\end{equation} 
With this decomposition, 
the eigenvectors of 
$\mathbb{S}$
mathematically correspond to spin up and spin down along the 
$\hat{\c B}$
axis, 
and are given by
\begin{equation}
|{+} \hat{\c B} \, \rangle
=
\begin{pmatrix}
\cos \frac{\Theta}{2} \\
- i e^{i \Phi} \sin \frac{\Theta}{2}
\end{pmatrix}
,\,\,\, 
|{-} \hat{\c B} \, \rangle
=
\begin{pmatrix}
\sin \frac{\Theta}{2}
\\
i e^{i \Phi} \cos \frac{\Theta}{2} 
\end{pmatrix}
\label{eq:UPDOWN}
.\end{equation}
These are column representations in the symmetric and antisymmetric basis; 
thus, 
there is an angle
$\Theta$
related to mixing,
and a relative phase
$\Phi$
between partial waves. 
These angles are determined by
\begin{eqnarray}
k \cot \Theta
&=&
-
\frac{\be + k^2 \d}{\sqrt{(\a - \g)^2 + 4 \sin^2 \phi\,}}
,\notag\\
\tan 
\Phi 
&=&  
- \frac{2 \sin \phi}{\a - \g} 
\label{eq:ANGLES}
,\end{eqnarray}
where the latter is time-reversal odd  and parity even. 
The non-standard convention employed for the azimuthal angle 
$\Phi$, 
moreover,
has the feature that 
$\phi = 0$
corresponds to 
$\Phi = 0 \, \text{mod} \, \p$, 
which is the requirement of a time-reversal even point interaction. 
For a parity-even point interaction, 
there is a similar requirement on the mixing angle, 
namely 
$\Theta = 0 \, \text{mod} \, \p$. 
Note that the quantity 
$k \cot \Theta$
is a linear function of the scattering energy
$k^2 = 2 m E$, 
and has a finite limit at threshold
$k = 0$.

Finally, 
we utilize all of these relations to parameterize the scattering 
$T$-matrix
Eq.~\eqref{eq:Tmatrix}. 
Employing the eigenstate scattering amplitudes 
$f_\pm$, 
which are defined by
\begin{equation}
f_\pm
= 
\frac{e^{2 i \d_\pm} - 1}{2i} 
= 
\frac{1}{\cot \d_\pm - i}
\label{eq:fpm}
,\end{equation}
the 
$T$-matrix in the partial-wave basis can be written in the form
\begin{equation}
\mathbb{T}
=
\begin{pmatrix}
\ol f  +  \D  f  \, \cos \Theta 
& 
\D f \, 
i e^{-i \Phi} 
\sin \Theta
\\
- \D f \,
i e^{ i \Phi} 
\sin \Theta
& 
\ol f  -  \D  f  \, \cos \Theta
\end{pmatrix}
\label{eq:T}
,\end{equation}
where
\begin{equation}
\ol f = \frac{1}{2} (f_+ + f_-),
\text{ and }
\D f = \frac{1}{2} ( f_+ - f_-)
.\end{equation}
This constitutes the full solution to the problem of scattering from a finite-range interaction. 
The phase shifts 
$\d_\pm(k)$
are determined from the eigenvectors of 
$\mathbb{T}$
via 
Eq.~\eqref{eq:fpm},
while the mixing angle and relative phase follow from the relations
\begin{equation}
e^{ 2 i \Phi} =  - \frac{\mathbb{T}_{10}}{\mathbb{T}_{01}}, 
\text{ and }
\tan^2 \Theta  = 
\frac{4 \mathbb{T}_{01} \mathbb{T}_{10}}{\left( \mathbb{T}_{00} - \mathbb{T}_{11} \right)^2}\
\label{eq:angles}
.\end{equation}
In turn, 
these quantities are related to the self-adjoint extension parameters of the general point interaction through 
Eqs.~\eqref{eq:EVPhases} and \eqref{eq:ANGLES}.

\subsection{Limiting Cases}                          %
\label{s:SAElimits}                                         %

Various limiting cases of Eq.~\eqref{eq:T} enable better understanding 
of the scattering matrix with a general point interaction. 
We consider parity-even interactions, 
time-reversal even interactions, 
and
$PT$-even interactions. 
Three special cases are also considered:
maximal time-reversal violation, 
decoupling of one pole,
and the case of scale-invariant point interactions.

\subsubsection{Parity-Even Interaction}     %
\label{s:PE}

A parity-even point interaction represents a important limiting case to detail fully. 
In this case, 
the self-adjoint extension parameters satisfy 
$\a = \g$
and
$\phi = 0$. 
Consequently,
$\Theta$
has two possible values
$\Theta = 0$ 
or 
$\p$. 
The second possibility corresponds to an inversion of 
$\hat{\c B}$
about the third direction, 
which would only result in a permutation of the eigenvectors in 
Eq.~\eqref{eq:UPDOWN}. 
Without loss of generality, 
we take 
$\Theta = 0$, 
so that the up (down) eigenvector corresponds to the symmetric (antisymmetric) partial wave. 
For a parity-even interaction,
$\Phi$
in Eq.~\eqref{eq:ANGLES}
becomes undefined without further specification of how the limits
$\g \to \a$
and
$\phi \to 0$
are taken. 
This is of no consequence, 
however, 
because the scattering matrix in Eq.~\eqref{eq:T} becomes independent of 
$\Phi$, 
namely
\begin{equation}
\mathbb{T}
= 
\text{diag} \big(  f_+, \, f_- \big)
.\end{equation}
From the 
$T$-matrix, 
we see that the partial-wave amplitudes are identical to the eigenstate amplitudes. 
We thus identify the phase shifts as those of the symmetric and antisymmetric partial waves
$\d_0 = \d_+$
and
$\d_1 = \d_-$. 
Note that for a parity-even interaction, 
one has 
$\k_\pm = (-\a \pm 1) /\d$
from 
Eq.~\eqref{eq:roots}.
Turning to 
Eq.~\eqref{eq:EVPhases}, 
we obtain the phase shifts in the form
\begin{equation}
e^{2 i \d_\pm} 
=
\pm \, \frac{k + i \k_\pm}{k - i \k_\pm}
\label{eq:PEphases}
,\end{equation}
which are appropriately unimodular.

The partial-wave scattering amplitudes can be expressed in terms of phase shifts in a way that exhibits their 
low-energy behavior.
For the symmetric amplitude, 
we write it in the form
\begin{equation}
f_0 = \frac{- i k \tan \d_0}{- k \tan \d_0 - i k} 
\label{eq:f0PE}
,\end{equation}
to expose that its poles in the complex momentum plane are determined by 
$- k \tan \d_0$. 
The low-energy behavior of this quantity is expected to have a well-behaved expansion
\begin{equation}
- k \tan \d_0 = - \frac{1}{a_0} + \frac{1}{2} r_0 \, k^2 + \c O(k^4)
\label{eq:ERE0}
,\end{equation}
in comparison with the three-dimensional effective-range expansion. 
For the parity-even point interaction
Eq.~\eqref{eq:PEphases}, 
we have
\begin{equation}
- k \tan \d_0 
=
- \k_+
,\end{equation}
which leads to the identification of the 
$\ell =0$ 
scattering length 
\begin{equation}
a_0 = \frac{\d}{1-\a}
,\end{equation}
in terms of the self-adjoint extension parameters.

The antisymmetric scattering amplitude, 
by contrast, 
is written in the form
\begin{equation}
f_1 = \frac{k}{k \cot \d_1 - i k}
\label{eq:f1PE}
,\end{equation}
to expose that its poles in the complex momentum plane are determined by 
$k \cot \d_1$, 
which is assumed to have the well-behaved low-energy limit 
\begin{equation}
k \cot \d_1 = - \frac{1}{a_1} + \frac{1}{2} r_1 \, k^2 + \c O(k^4)
\label{eq:ERE1}
.\end{equation}
The 
$\ell = 1$
effective-range expansion in one dimension is of the same form as the expansion for
$\ell = 0$
in three dimensions.%
\footnote{
For reference, 
behavior of the
$\ell = 1$
scattering amplitude in three dimensions is exhibited by writing
$f_1(k) = \frac{k^2}{k^3 \cot \d_1 - i k^3}$, 
and noting that the quantity 
$k^3 \cot \d_1 = - \frac{1}{(a_1)^3} + \frac{1}{2 r_1} \, k^2 + \c O(k^4)$
is amenable to a low-energy expansion%
~\cite{taylor2006scattering}. 
} 
For the parity-even point interaction 
Eq.~\eqref{eq:PEphases}, 
we have
\begin{equation}
k \cot \d_1 
=
- \k_-
,\end{equation}
which leads to the identification of the 
$\ell = 1$
scattering length
\begin{equation}
a_1 = - \frac{\d}{1+\a}
.\end{equation}
In the case of a parity-even point interaction, 
the two independent self-adjoint extension parameters 
$\a$
and
$\d$
thus determine the scattering lengths of the uncoupled
$s$- 
and 
$p$-waves.

\subsubsection{Time-Reversal--Even Interaction}%
\label{s:TREI}                                                         %

In the case of a point interaction that is time-reversal even, 
the self-adjoint extension parameter 
$\phi$
vanishes, 
leaving three unconstrained parameters.
As a consequence, 
the phase 
$\Phi$
in Eq.~\eqref{eq:ANGLES}
has the value
$\Phi = 0 \, \text{mod} \, \p$, 
so that
$\cos \Phi = \pm 1$,
with the sign determined by 
$\sgn(\a - \g) = \pm 1$. 
As parity is generally still broken,
there is mixing between partial waves;
and, 
the mixing angle becomes 
$\Theta$,
up to a constant of proportionality. 
For simplicity, 
we consider the case 
$\Phi = 0$;
the formulas for 
$\Phi = \p$
differ only by the sign of 
$\Theta$, 
which could be absorbed by redefining the mixing angle.

The partial-wave  
$T$-matrix in 
Eq.~\eqref{eq:T}
can be used to find the scattering amplitude for an incoming right-traveling wave to be found as a symmetric outgoing wave
\begin{equation}
f_0^{(+)}
=
e^{ \frac{i \Theta}{2}}
\big[
f_+ \, \cos \tfrac{\Theta}{2} - i f_- \, \sin \tfrac{\Theta}{2}
\big]
\label{eq:f0RPV}
,\end{equation}
as well as an antisymmetric outgoing wave
\begin{equation}
f_1^{(+)}
=
e^{ \frac{i \Theta}{2}}
\big[
f_- \, \cos \tfrac{\Theta}{2} - i f_+ \, \sin \tfrac{\Theta}{2}
\big]
\label{eq:f1RPV}
.\end{equation}
For each of these partial-wave amplitudes, 
the modulus-squared coefficients sum to unity,
and the mixing angle can be identified as 
$\frac{\Theta}{2}$.  
If one is interested solely with the scattering of an incoming right-traveling wave, 
the identical overall phases are irrelevant. 
One must be careful, 
however, 
because the scattering amplitudes for an incoming left-traveling wave are related by a parity transformation to those above, 
namely
\begin{equation}
f_\ell^{(-)} = f_\ell^{(+)} \Big|_{\Theta \to - \Theta}
\label{eq:PPWs}
.\end{equation}

\subsubsection{PT-Even Interaction}%

In the case of a 
$PT$-even interaction, 
we have 
$\a = \g$
for any value of the 
$PT$-symmetric parameter
$\phi$. 
As a consequence, 
the relative phase satisfies
$\Phi = \mp \frac{\p}{2}$, 
the value of which depends on 
$\sgn(\phi) = \pm 1$. 
For simplicity, 
we consider only the case 
$\Phi = - \frac{\p}{2}$; 
the formulas for 
$\Phi = \frac{\p}{2}$
differ only by the sign of 
$\Theta$, 
which could be absorbed by merely redefining this angle. 
Mixing in this case is solely due to parity violation introduced by 
$\phi$.
Using the 
$T$-matrix 
in the partial-wave basis  
Eq.~\eqref{eq:T}, 
the scattering amplitude for an incoming right-traveling wave to be found as a symmetric outgoing wave is
\begin{multline}
f_0^{(+)}
=
\cos \tfrac{\Theta}{2} 
\Big[
f_+ \,  \cos \tfrac{\Theta}{2} \left( 1  - \tan \tfrac{\Theta}{2} \right)  
\\
+ 
f_- \, \sin \tfrac{\Theta}{2} \left( 1 + \tan \tfrac{\Theta}{2}  \right)
 \Big]
\label{eq:f0PT}
,\end{multline}
while that for an antisymmetric outgoing wave is
\begin{multline}
f_1^{(+)}
=
\cos \tfrac{\Theta}{2}
\Big[ 
f_-  \, \cos \tfrac{\Theta}{2}  \left(  1 + \tan \tfrac{\Theta}{2} \right)
\\
- 
f_+ \, \sin \tfrac{\Theta}{2} \left( 1 - \tan \tfrac{\Theta}{2} \right)
\Big]
\label{eq:f1PT}
.\end{multline}
The partial-wave scattering amplitudes for an incoming left-traveling wave are related by the parity transformation
in 
Eq.~\eqref{eq:PPWs}.

There is a complicated feature of the mixing present in this case 
that is also shared by the general case. 
The modulus-squared coefficients of the two eigenstate scattering amplitudes do not sum to unity 
in each partial wave. 
Instead, 
the sum of all four (over both partial waves) is unity, 
which is consistent with probability conservation. 
There is conventional mixing between the eigenstate basis and the partial-wave basis, 
not between the right- and left-traveling basis and the partial-wave basis. 
For example, 
the 
$f_-$
amplitude in 
Eqs.~\eqref{eq:f0PT} and \eqref{eq:f1PT}
has the same overall factor 
$\cos \frac{\Theta}{2} \left(1 + \tan \frac{\Theta}{2} \right)$, 
but is additionally accompanied by 
$\sin \frac{\Theta}{2}$
in the symmetric wave, 
and
$\cos \frac{\Theta}{2}$
in the antisymmetric wave. 
The down eigenstate thus mixes with an angle of 
$\frac{\Theta}{2}$
into the partial-wave basis. 
This more intricate pattern of mixing occurs when time-reversal is broken.

\subsubsection{Maximal Time-Reversal Violation}%

There is curious behavior in the case of maximal
time-reversal violation, 
which is attained when  
$\phi = \frac{\pi}{2}$. 
Larger values of 
$\phi$ 
are redundant because the range of
$\phi$ can be reduced to 
$(-\frac{\p}{2},\frac{\p}{2}\,]$
by adjusting the signs of the other self-adjoint extension parameters. 
In the case of maximal time-reversal violation, 
the eigenvalues of the 
$S$-matrix become
\begin{equation}
e^{2 i \d_\pm}
=
\pm \sqrt{\frac{(k+i\k_+)(k+i\k_-)}{(k-i\k_+)(k-i\k_-)}}
\label{eq:Swhat}
.\end{equation}
Strikingly, 
there are no longer poles of the 
$S$-matrix on the imaginary momentum axis;
instead, 
the former pole locations have become branch points.
There is no conflict with unitarity, 
as each eigenvalue of 
$\mathbb{S}$
is manifestly unimodular. 
For all momenta
$k$, 
furthermore,
the eigenstate phase shifts always differ in phase by 
$\frac{\p}{2}$.

Supposing that
$\k_+$
is positive, 
for example,
there will be a bound state.%
\footnote{
In the case of negative values, 
there will be antibound states
(also called virtual states or virtual bound states). 
While such non-normalizable solutions do not correspond to physical states, 
they too show up as poles of the 
$S$-matrix
and can have a sizable effect on low-energy scattering near threshold. 
The $S$-matrix in Eq.~\eqref{eq:Swhat}, 
however, 
does not exhibit any poles. 
}
For algebraic simplicity, 
we take 
$\a = \g$
in what follows, 
but note that adopting $PT$ symmetry is not necessary. 
Up to normalization, 
the coordinate wavefunction of the bound state can be obtained
by taking the residue of the incoming right- or left-traveling solutions in 
Eq.~\eqref{eq:Rinc}
at 
$k = i \k_+$. 
Taking residues of both solutions leads to the same bound-state wavefunction
\begin{equation}
\psi_E(x)  
=
\sqrt{\tfrac{\k_+}{2}}
\, e^{ - \k_+ |x|}
\big[
1
+
i \sgn(x)
\big]
,\end{equation}
up to an overall phase.
With a real phase convention chosen, 
the wavefunction is invariant under the $PT$ transformation
$\left[ \psi_E(-x) \right]^* = \psi_E(x)$. 
This normalized wavefunction satisfies the boundary conditions in 
Eq.~\eqref{eq:SAE1D},
and has the bound-state energy 
$E = - \frac{\k_+^2}{2m}$. 
Maximal time-reversal violation, 
however, 
seems to be a case for which bound states do not appear 
as poles of the $S$-matrix.

\subsubsection{One Pole Decouples}                 %
\label{sec:onepole}                                              %

Each 
$S$-matrix 
eigenvalue in 
Eq.~\eqref{eq:EVPhases}
generally exhibits one pole.
Specifically, 
$\k_\pm$
is the pole location of the eigenvalue
$e^{2 i \d_\pm}$.%
\footnote{
To analytically continue the numerator to 
$k = i \k_\pm$, 
we choose the branch of 
$| \vec{\mathcal{B}} \, |$ 
so that the eigenvalues are continuously connected to those 
in the parity-even limit 
Eq.~\eqref{eq:PEphases}.
This requires 
$| \vec{\mathcal{B}} \, | \to \mp 2 \k_\pm \cos \phi$,
and establishes that each $S$-matrix eigenvalue has only one pole.
} 
When 
$\k_\pm > 0$, 
the state will be a bound state in the spectrum of 
$H$; 
while, 
for 
$\k_\pm < 0$, 
the pole is on the unphysical sheet 
and corresponds to an antibound state. 
Only for a parity-even point interaction is 
each 
$S$-matrix pole associated with 
a symmetric or antisymmetric state.  
When parity is broken, 
each state is a superposition having indefinite parity.

A possible scenario at low energies is that one pole is closer to threshold, 
and the other decouples. 
Returning to 
Eq.~\eqref{eq:Den}, 
this will be the case when either
$\d = 0$
or 
$\be = 0$.%
\footnote{ 
When both vanish, 
there are no poles.
This case is that of the scale-invariant point interaction addressed in
Sec.~\ref{s:SAEscale}. 
}
Due to the constraint in 
Eq.~\eqref{eq:det}, 
one must have 
$\g = \a^{-1}$
for both of these possibilities.

For the first possibility, 
we restrict to 
$\d = 0$, 
for which
the denominator 
$\c D$
has one root located at 
$k = i \k_0$, 
where
\begin{equation}
\k_0 = - \frac{\be}{\a + \a^{-1}} 
.\end{equation}
Consequently, 
the eigenstate scattering amplitudes become
\begin{equation}
f_\pm
{=}
\frac{2 i k \cos \phi {+} i \frac{\be}{\k_0} ( k {-} i \k_0 ) {\mp} \sqrt{\frac{\be^2}{\k_0^2} (k^2  {+} \k_0^2) {-} 4 k^2 \cos^2 \phi \,}}
{2 \frac{\be}{\k_0} ( k - i \k_0 )}
\label{eq:fpmd=0}
,\end{equation}
but only 
$f_+$
has a pole at 
$k = i \k_0$. 
Due to breaking of parity and time reversal,
the mixing angle and relative phase are generally non-vanishing
\begin{eqnarray}
k \cot \Theta
&=&
-
\frac{\be}{\sqrt{(\a - \a^{-1})^2 + 4 \sin^2 \phi \,}}
,\notag \\
\tan 
\Phi 
&=&  
- \frac{2 \sin \phi}{\, \a - \a^{-1}} 
\label{eq:ANGLES0}
.\end{eqnarray}
Note that the pole location alone does not fix these quantities. 
The energy of the bound-state (or antibound-state) pole
is necessarily a parity even and time-reversal even quantity. 
The scattering matrix is required to glean information about
the breaking of parity and time-reversal. 
If one takes the limit of parity and time-reversal invariance, 
$\Theta \to 0$
along with
$f_- \to 0$,
and only then is the pole exclusively an 
$s$-wave.

For the other possibility, 
we restrict to 
$\be = 0$, 
for which there will be a pole at 
$k = i \k_1$, 
where
\begin{equation}
\k_1
= - \frac{\a + \a^{-1}}{\d}
.\end{equation}
Note that while 
$\c D$
has a root at threshold, 
the eigenstate scattering amplitudes remain finite at 
$k = 0$.
These amplitudes have the form
\begin{equation}
f_\pm
{=}
\frac{2 i \cos \phi - \d ( k - i \k_1 ) \pm \sqrt{\d^2 (k^2  + \k_1^2) {-} 4 \cos^2 \phi \,}}
{2 i \d ( k - i \k_1 )}
\label{eq:fpmb=0}
,\end{equation}
and 
$f_-$
exhibits the pole. 
When one takes the limit of parity and time-reversal invariance, 
then
$\Theta \to 0$
along with
$f_+ \to 0$, 
and the pole in 
$f_-$ 
becomes exclusively 
$p$-wave. 
The mixing angle and relative phase in the
$\be = 0$
case are given by
\begin{eqnarray}
k \tan \Theta
&=&
-
\frac{\sqrt{(\a-\a^{-1})^2 + 4 \sin^2 \phi \,}}{\d}
,\notag\\
\tan 
\Phi 
&=&  
- \frac{2 \sin \phi}{\, \a - \a^{-1}} 
\label{eq:ANGLES1}
.\end{eqnarray}
The two possibilities
$\d = 0$
and
$\be = 0$ 
can be distinguished, 
for example, 
by the markedly different behavior of the mixing angle near threshold.

\subsubsection{Scale-Invariant Interaction}
\label{s:SAEscale}

For a scale-invariant point interaction,%
\footnote{For a scale-invariant point interaction, 
the scaling property of the Hamiltonian
$H \to \lambda^{-2} \, H$
cannot be modified by the self-adjoint extension to 
$x = 0$. 
A scale-invariant interaction thus preserves the scale transformation of the Hamiltonian.  
} 
both dimensionful parameters vanish
$\be = \d = 0$,
and the 
$S$-matrix does not have poles on the imaginary momentum axis. 
This can be argued on physical grounds, 
because a pole would imply the existence of an energy scale. 
To obtain the scattering matrix for this special case, 
we take the 
$\d \to 0$
limit of the 
$\be = 0$
result in
Eq.~\eqref{eq:fpmb=0}. 
With a scale-invariant point interaction, 
the eigenstate scattering amplitudes have the form 
\begin{equation}
f_\pm
=
\frac{2 \cos \phi - (\a + \a^{-1}) \pm \sqrt{4 \cos^2 \phi  {-} (\a + \a^{-1})^2 \,}}
{2 i  (\a + \a^{-1}) }
\label{eq:fpmb=d=0}
,\end{equation}
which are momentum independent and anomalous at threshold. 
The formula for the relative phase is unchanged from 
Eq.~\eqref{eq:ANGLES1}, 
while the mixing angle curiously satisfies
$\frac{\Theta}{2} = \sgn(\a) \, \frac{\p}{4}$,
for all momenta. 

To make the connection with Levinson's theorem in one dimension, 
one requires a parity-even interaction. 
This imposes 
$\phi = 0$
and
$\a = \pm 1$,
for all values of 
$k$. 
Consequently, 
results agree with Levinson's theorem
as
$\d_\ell (0) = 0$ 
or 
$\frac{\p}{2}$
depending on 
$\sgn(\a)$. 
The former result is that of a free particle, 
while the latter result merely reflects the overall reversal of sign at the origin; 
these cases are trivial point interactions. 
Classical scale symmetry is present in the low-energy effective theory
Sec.~\ref{s:C1D} only for two possible contact interactions that are
however,
parity odd.

\section{Contact Interactions in One Dimension}
\label{s:C1D}

The general point interaction must be describable in terms of a low-energy effective theory of contact interactions. 
The  zero-range effective interaction 
$V$
in one dimension has the general form 
\begin{equation}
V =  - \frac{1}{m} \sum_{j=0}^\infty \c O_j
\, ,\end{equation}
and contains infinitely many contributions that are indexed by 
$j$, 
which is the number of derivatives
(momentum operators)
appearing in the various terms of each contribution
$\c O_j$. 
Such terms are restricted by Hermiticity, 
and our interest is only with those that are additionally self adjoint. 
Note that for algebraic convenience, 
we have factored out 
$-m^{-1}$
from all contributions. 
Operator coefficients are real valued in what follows. 
These are generally running couplings, 
and contributions from the corresponding terms of 
$\c O_j$
are regulated. 
The regularization schemes we employ are detailed in \ref{s:DimReg}.

To discuss properties of contact operators 
with respect to the partial-wave basis, 
we use the parity operator 
$\c P$, 
which satisfies 
$\c P^2 = 1$. 
In the partial-wave basis, 
the symmetric and antisymmetric waves are even- and odd-parity eigenstates, 
so that
$\c P \, | \ell \rangle = (-1)^\ell \, | \ell \rangle$. 
Positive and negative parity projection operators are defined by 
$\c P_\pm = \frac{1}{2} \left( 1 \pm \c P \right)$, 
and accordingly satisfy the relations 
$\c P_\pm^2 = \c P_\pm$
and 
$\c P_\pm \, \c P_\mp = 0$.
In one dimension, 
the operator 
$\c P_+$
projects onto $s$-waves, 
while 
$\c P_-$
projects onto $p$-waves. 
The parity-odd transformations of the position and momentum operators are encoded in the anticommutation relations
$\{ \c P, x \} = \{ \c P, p \} = 0$. 
Consequently, 
parity-odd operators, 
such as the position operator
$x$,
have the property 
$x \, \c P_\pm = \c P_\mp \, x$.

With zero derivatives, 
the contribution 
$\c O_0$
has only one term
\begin{equation}
\c O_0 = c_0 \, \delta(x)
,\end{equation}
which is the Dirac delta-function interaction. 
This term has even parity
$[ \, \c P, \d (x) \,] = 0$, 
which leads to the decomposition 
\begin{equation}
\d(x)
= 
\c P_+ \, \d(x) \, \c P_+
+
\c P_- \, \d(x) \, \c P_-
\longrightarrow
\c P_+ \, \d(x) \, \c P_+
.\end{equation} 
Note that the second term in this decomposition can be dropped. 
It couples only to antisymmetric waves, 
and these vanish at the origin
$\langle \ell{=}1 | \, \d(x) \, | \ell{=} 1 \rangle = 0$.

For the contribution 
$\c O_1$, 
there are two terms with one derivative 
\begin{equation}
\c O_1 = c_1 \, i \left[ \, p, \delta(x) \, \right] + \widetilde{c}_1 \, \left\{ p, \delta(x) \right\}
\label{eq:O1}
.\end{equation}
The first term 
$i \left[ \, p, \delta(x) \, \right] = \delta'(x)$
is simply the derivative of the delta-function interaction. 
Both terms of 
$\c O_1$
have odd parity,
and they mediate transitions between symmetric and antisymmetric waves, 
which is manifest in the identity
$\c O_1 = \c P_+ \, \c O_1 \, \c P_- + \c P_- \, \c O_1 \, \c P_+$. 
The term with coefficient 
$c_1$
is 
time-reversal even, 
while the term with coefficient
$\widetilde{c}_1$
is time-reversal odd,
hence $PT$ even. 
Under scaling
$x \to \lambda \, x$, 
these terms have the transformation
$\c O_1 \to \lambda^{-2} \, \c O_1$, 
using that 
$p \to \lambda^{-1} \, p$. 
Excluding all other contributions to  
$V$, 
the Hamiltonian 
$H = \frac{p^2}{2m} - \frac{1}{m} \c O_1$
has the scale transformation
$H \to \lambda^{-2} \, H$. 
In this respect, 
the two terms in 
$\c O_1$
are unique.

Terms contributing at the next order have two derivatives, 
which can be systematically generated from the three building blocks
$p^2 \d(x)$, $\d(x) \, p^2$, and $p \, \d(x) \, p$. 
Two Hermitian combinations can be formed from the first two
$\big\{ p^2, \d(x) \big\}$ 
and 
$i \big[ p^2, \d(x) \big]$, 
while the third building block is already Hermitian. 
Thus, 
we choose the basis of second-derivative operators 
\begin{equation}
\c O_2 
= 
c_2^{(p)}
\, p \, \d(x) \, p
+
c_2 \, \big\{ p^2, \d(x) \big\}
+ 
\widetilde{c}_2 \,
i \big[ p^2, \d(x) \big] 
\label{eq:2Ds}
.\end{equation}
The first two terms of 
$\c O_2$
are even under time reversal, 
while the last term is odd. 
All three terms are even under parity, 
which is made manifest by the identity 
$\c O_2 = \c P_+ \, \c O_2 \, \c P_+ + \c P_- \, \c O_2 \, \c P_-$. 
As these operators involve contact interactions, 
moreover, 
they exhibit further selectivity in the partial-wave basis. 
Using the property that antisymmetric waves vanish at the origin, 
we observe that
\begin{multline}
\c O_2 
= 
c_2^{(p)} \, 
\c P_- \, p \, \d(x) \, p \, \c P_-
+
c_2 \, \c P_+ \, \big\{ p^2, \d(x) \big\} \, \c P_+
\\
+ 
\widetilde{c}_2 \, 
\c P_+ \,
i \big[ p^2, \d(x) \big] \, \c P_+
\label{eq:pwas}
.\end{multline}
The operator with coefficient 
$c_2^{(p)}$
couples only to
$p$-waves, 
while the operators with coefficients 
$c_2$
and
$\widetilde{c}_2$
couple only to 
$s$-waves.

Any additional second-derivative terms are redundant due to relations that exist between operators. 
For example, 
the operator 
$\big\{ p, \d'(x) \big\}$
is already accounted for in Eq.~\eqref{eq:2Ds},
due to the relation 
$\big\{ p, \d'(x) \big\} = i \big[ p^2, \d(x) \big]$.
Concerning the 
$\d''(x)$
operator, 
note that the identity
\begin{equation}
\d''(x) 
= 
- \big[ \, p, \left[ \, p, \d(x)\, \right] \, \big] 
= 
2 p \, \d(x) \, p - \big\{ p^2, \d(x) \big\}
\label{eq:relation}
,\end{equation} 
establishes its linear dependence. 
From 
Eq.~\eqref{eq:pwas}, 
moreover, 
we see that 
$\d''(x)$
contributes to interactions of both
$s$- and $p$-waves.

\begin{table}[tb]
\begin{center}
\begin{tabular}{||c||c|c|c|c|c||}
\hline
\hline
Operator & $\ell$  
& $\  P \ $ & $T$ & Scale & Adj \\
\hline
$c_0 \, \d(x)$ & $s$ & $+$ & $+$ & $\lambda^{-1}$ & \ding{51} \\
$c_1 \,  i \big[ p, \d(x) \big]$ & $ s {\leftrightarrow} p$ & $-$ & $+$ & $\lambda^{-2}$ & \ding{51} \\
$\widetilde{c}_1 \, \left\{ p, \d(x) \right\}$ & $ s {\leftrightarrow} p$ & $-$ & $-$ & $\lambda^{-2}$ & \ding{51} \\
$c^{(p)}_2 \, p \, \delta(x) \, p$ & $p$ & $+$ & $+$ & $\lambda^{-3}$ & \ding{51} \\
\hline
\hline
$c_2 \, \big\{ p^2, \d(x) \big\}$ & $s$ & $+$ & $+$ & $\lambda^{-3}$ & \ding{55} \\
$\widetilde{c}_2 \, i \big[ p^2, \d(x) \big]$ & $s$ & $+$ & $-$ & $\lambda^{-3}$ & \ding{55} \\
\hline
\hline
\end{tabular}
\end{center}
\caption{Contact interactions up to second-derivative order, and their symmetry properties.
Relevant partial waves are denoted by 
$\ell$, 
while parity 
$P$
and time-reversal 
$T$
properties are also listed
($+$ 
for even, 
$-$ 
for odd).
Scale refers to the operator's transformation under the rescaling
$x \to \lambda \, x$.
The operator with coefficient 
$c_1$
is simply 
$\d'(x)$, 
while the operator with coefficient 
$c_2$
is the linear combination of 
$\d''(x)$
and 
$p \, \d(x) \, p$
shown in 
Eq.~\eqref{eq:relation}.
The last column lists whether the operator is self adjoint, 
and computations are carried out only for the self-adjoint operators.
}
\label{t:Ops}
\end{table}

Up to second-derivative order, 
a total of six contact interactions appear in the effective Hamiltonian. 
These are collected in Table~\ref{t:Ops}, 
but not all are self adjoint. 
Once regulated, 
the second-derivative operators with coefficients 
$c_2$
and
$\widetilde{c}_2$
correspond to energy-dependent point interactions. 
Accordingly, 
these operators are omitted in the calculations that follow,%
\footnote{
The operator with coefficient 
$\widetilde{c}_2$
can be ruled out based on symmetry, 
because there is no parity-even point interaction that is time-reversal odd. 
Additionally due to its structure, 
the operator cannot modify the renormalized scattering amplitudes. 
In field theory parlance, 
it constitutes an equation-of-motion operator, 
and such operators are redundant. 
Using the effective Hamiltonian
$H$, 
we can write
$[ \, p^2, \d(x) \, ] = 2 m \big[ H - E, \, \d(x) \, \big]  {+} \text{ singular}$,
where
$E$
is the energy eigenvalue, 
and singular represents terms that have a product of two delta functions. 
Such singular contributions will always be absorbed by a renormalization condition. 
After renormalization, 
the operator with coefficient 
$\widetilde{c}_2$
will thus make a vanishing contribution to any eigenstate matrix elements.
The redundancy of equation-of-motion operators in quantum field theory holds on shell
(for eigenstates)
as well as off shell
(for Green's functions)%
~\cite{Arzt:1993gz}. 
The commutator identity also shows that the renormalized contributions from this operator 
also vanish when acting between time-independent Green's functions. 
This is due to the defining equation 
$\left( H - E \right) G_E(x,0) = \delta(x)$,
and the fact that 
$\d(x)$
commutes with itself. 
}
leaving four self-adjoint contact interactions.

\subsection{Momentum-Space Scattering Solutions}

Given the momentum dependence of the contact interactions, 
it is natural to reformulate solutions to the scattering problem directly in momentum space. 
Employing the Fourier transform of the wavefunction
\begin{equation}
\phi(p)
=
\int_{-\infty}^{+\infty} dx \, e^{- i p x} \psi(x)
,\end{equation}
we can write the momentum-space solutions for incoming symmetric and antisymmetric 
waves scattering off contact interactions as%
\footnote{
There is an additional term 
needed in the most general solution, 
which takes the form of a polynomial in momentum
$\phi_{\ell, \text{short}}(p) =  \sum_{n=0}^\infty \, g_{n \, \ell} \, (i p)^n$. 
In coordinate space, 
this polynomial generates short-distance contributions to the wavefunction 
$\psi_{\ell, \text{short}}(x) = \sum_{n=0}^\infty \, g_{n \, \ell} \, \d^{(n)}(x)$, 
which ultimately require a regularization scheme to handle consistently. 
While such contributions are absent from the asymptotic scattering states, 
they are required to obtain solutions for the $T$-matrix elements. 
In the calculations that follow, 
however,  
the coefficients
$g_{n \, \ell}  = 0$, 
because only energy-dependent contact interactions can produce 
$\phi_{\ell, \text{short}}(p)$.
For this reason, 
the short-distance contribution is omitted from 
Eq.~\eqref{eq:phi}. 
}
\begin{equation}
\phi_\ell(p)
=
\phi_{\ell,\text{inc}}(p) + \phi_{\ell, \text{out}}(p) 
\label{eq:phi}
,\end{equation}
where the solutions are labeled by 
$\ell = 0$ 
($\ell = 1$) 
for the symmetric (antisymmetric) incoming waves. 
In momentum space, 
delta functions describe the incoming waves
\begin{equation}
\phi_{\ell, \text{inc}}(p) 
= 
2\p \, 
\frac{\d(p - k) {+} (-1)^\ell \d(p+k)}{2} 
,\end{equation}
where 
$k$ 
is the magnitude of the scattering momentum. 
The incoming waves are solutions to the homogeneous momentum-space Schr\"odinger equation 
$(p^2 - k^2) \phi_{\ell, \text{inc}}(p) = 0$.
The outgoing momentum-space wavefunction has the form
\begin{equation}
\phi_{\ell, \text{out}} (p) 
= 
2 \,
\frac{k \, \mathbb{T}_{0\,\ell} + p \, \mathbb{T}_{1\,\ell}}{p^2 - k^2 - i \epsilon} 
\label{eq:phiout}
.\end{equation}

The inverse Fourier transform of 
$\phi_\ell(p)$ 
exposes the significance of terms appearing in the decomposition of the momentum-space solution.
The corresponding coordinate wavefunction is
\begin{multline}
\psi_\ell(x)
=
i^\ell 
\cos \left( k x - \tfrac{\ell \p}{2} \right)
\\
+
i \, e^{i k |x|}
\Big[ 
\mathbb{T}_{0 \, \ell} 
+  
\sgn(x) \,\mathbb{T}_{1 \, \ell}
\Big]
\label{eq:psi}
.\end{multline}
The first term arises from the sum and difference of momentum-space delta-functions, 
which produce the incoming symmetric and antisymmetric waves, 
respectively. 
Terms of the momentum-space wavefunction 
$\phi_{\ell, \text{out}}(p)$
with amplitudes
$\mathbb{T}_{0 \, \ell}$
and
$\mathbb{T}_{1 \, \ell}$
produce the outgoing symmetric and antisymmetric waves 
in the form of  
Eq.~\eqref{eq:partialpartial}. 
These amplitudes are 
$T$-matrix elements in the partial-wave basis, 
see 
Eq.~\eqref{eq:Tmatrix}.

With contact interactions, 
the momentum-space Schr\"odinger equation 
\begin{equation}
(p^2 - k^2) \phi_\ell(p) 
= 
- 2 m \int_{-\infty}^\infty dx \, e^{-  i p x} \, V(x) \, \psi(x)
\label{eq:pSchro}
,\end{equation} 
leads to an expression for 
$\phi_\ell(p)$
in terms of regulated values of the coordinate wavefunction and its derivatives at the origin
$\psi_\ell^{(n)}(0)$.
From Eq.~\eqref{eq:phiout}, 
one identifies the 
$T$-matrix elements in terms of these origin values. 
Equations for the regulated origin values are then obtained by taking 
moments of the momentum-space wavefunction in
Eq.~\eqref{eq:phi}, 
namely
\begin{equation}
\psi^{(n)}_\ell(0) = \int \frac{dp}{2 \p} (i p)^n \phi_\ell(p)
\label{eq:moments}
.\end{equation}
This procedure results in a system of linear equations for the
$\psi^{(n)}_\ell(0)$, 
the algebraic solution of which enables determination of 
$\mathbb{T}$. 
In evaluating the wavefunction 
\eqref{eq:psi}
at 
$x=0$,
we are confronted with 
$\sgn(0)$, 
which is regularization scheme dependent. 
Consistency with  
Eq.~\eqref{eq:phiout}
requires
\begin{equation}
i \, \sgn(0)
=
\int \frac{dp}{2\p} \,
\frac{2 p}{p^2 - k^2 - i \epsilon} 
\equiv 0
,\end{equation} 
which is essentially zero by definition in any parity-preserving regularization scheme, 
including those that we employ.%
\footnote{
The scheme-dependent value 
$\sgn(0) = 0$
implies that a discontinuous function of the form
$f(x) = f_0(x) + \sgn(x) \, f_1(x)$ 
has the origin value
$f(0) = f_0(0) = \frac{1}{2} \left[ f(\e) + f(-\e) \right]$, 
which is identical to its average across the discontinuity. 
The derivative of this function
$f'(x) = f'_0(x) + \sgn(x) \, f'_1(x) + 2 \, \d(x) f_1(0)$ 
has an additional  
$\d(0)$
contribution at the origin that requires regularization. 
The identification of origin values as averages across the discontinuity proposed in
Ref.~\cite{Griffiths:1993}
omitted the regularization scale and scheme dependence, 
but was later reconsidered with a counterexample%
~\cite{10.1119/1.19283}. 
Na\"ive dimensional regularization actually provides a scheme in which all of the results of 
Ref.~\cite{Griffiths:1993}
hold and likely underlies the theory of discontinuous distributions of 
Ref.~\cite{KURASOV1996297};
however, 
the NDR scheme is not necessarily appropriate for every problem.
}
Note that derivatives of the coordinate wavefunction at the origin
\eqref{eq:moments}
require the regulator-dependent values of 
$\d^{(n)}(0)$
discussed in 
\ref{s:DimReg}.

In light of these technical points, 
we can obtain expressions for the origin values
$\psi_\ell^{(n)}(0)$
using the general form of the momentum-space scattering solution 
Eq.~\eqref{eq:phi}. 
With the set of energy-independent contact interactions, 
only the lowest two moments of the momentum wavefunction are required. 
Performing the momentum integrals in 
Eq.~\eqref{eq:moments}
results in the relations
\begin{align}
\psi_\ell(0) 
&=
\d_{0 \, \ell} + i \, \mathbb{T}_{0 \, \ell} 
,\notag \\
\psi^\prime_\ell(0)
&=
i k \, \d_{1 \, \ell}  + 2 i \, \mathbb{T}_{1 \, \ell} \ I_2(k)
\label{eq:psizeros}
.\end{align}
The integral 
$I_{2}(k)$
is defined and regulated in%
~\ref{s:DimReg}. 
The relations in 
Eq.~\eqref{eq:psizeros}
are employed below to obtain the scattering matrix for contact interactions.

\subsection{Parity-Even Contact Interactions}
\label{s:peci}

To illustrate how the procedure works, 
we begin by computing the $T$-matrix for scattering mediated by the two energy-independent, 
parity-even contact interactions in 
Table~\ref{t:Ops}. 
The computation is straightforward to perform using a hard momentum 
cutoff
$\L$, 
and we show that equivalent renormalized results are obtained with the NDR scheme. 
Scattering amplitudes are compared to those of the parity-even point interaction detailed in 
Sec.~\ref{s:PE}.

With these two parity-even contact interactions, 
the momentum-space Schr\"odinger equation~\eqref{eq:pSchro}
produces
\begin{multline}
\phi_\ell(p)
=
\phi_{\ell, \text{inc}}(p)
+ 2 \, 
\frac{
c_0 
\psi_\ell(0) 
- i p \, c_2^{(p)} \psi_\ell^\prime(0)}{p^2 - k^2 - i \epsilon}
\label{eq:phiPE}
.\end{multline}
Comparing with the general solution given in 
Eq.~\eqref{eq:phi}, 
we readily identify the partial-wave $T$-matrix elements
\begin{eqnarray}
k \, \mathbb{T}_{0 \, \ell} 
&=& 
\phantom{-} c_0 
\psi_\ell(0) 
,\notag \\
\mathbb{T}_{1 \, \ell} 
&=& 
- i c_2^{(p)} \psi^\prime_\ell(0)
.\end{eqnarray}
These quantities depend on values of the wavefunction and its derivative at the origin, 
which from 
Eq.~\eqref{eq:psizeros},
in turn, 
depend on the 
unknowns
$\mathbb{T}_{0 \, \ell}$, 
$\mathbb{T}_{1 \, \ell}$. 
Eliminating the origin values of the wavefunction and its derivative, 
we obtain the uncoupled equations
\begin{align}
k \, \mathbb{T}_{0 \, \ell} 
&=
\phantom{-} c_0 
 \left[ \d_{0 \, \ell} + i \, \mathbb{T}_{0 \, \ell} 
\right]
,\notag \\
\mathbb{T}_{1 \, \ell} 
&=
- i c_2^{(p)}  \big[ i k \, \d_{1 \, \ell} + 2 i \, \mathbb{T}_{1 \, \ell} \, I_2(k)  \big]
,\end{align}
for each value of
$\ell$. 
Using either a hard momentum cutoff 
$\L$
or in NDR, 
the function 
$I_2 (k)$ 
is replaced with its regulated value
$\frac{ik}{2} + \d(0)$, 
see 
Eqs.~\eqref{eq:I2L} and \eqref{eq:I2NDR}.

In solving the above equations, 
notice that the 
$T$-matrix has the form
$\mathbb{T} = \text{diag} \left( \mathbb{T}_{00}, \mathbb{T}_{11} \right)$, 
from which the diagonal elements are the partial-wave scattering amplitudes 
$f_0$ and $f_1$ of 
Eq.~\eqref{eq:Tpwa}, 
respectively. 
These amplitudes are found to be
\begin{equation}
f_0(k) 
{=} 
\frac{-i c_0}{ - c_0 - i k}
\label{eq:swave}
\end{equation}
for the $s$-wave, and
\begin{equation}
f_1(k)
= 
\frac{k}{\big[c_2^{(p)}\big]^{-1} {-} 2 \d(0)  -  ik}
\label{eq:pwave}
,\end{equation}
for the $p$-wave. 
While the latter requires renormalization, 
the former is the scale- and scheme-independent result
\begin{equation}
- k \tan \d_0 = - c_0
,\end{equation}
obtained by comparing 
$f_0(k)$
with 
Eq.~\eqref{eq:f0PE}.
This comparison enables identification of the coefficient 
$c_0 = (a_0)^{-1}$
in terms of the  
$s$-wave scattering length from 
Eq.~\eqref{eq:ERE0}.

To renormalize the $p$-wave amplitude
Eq.~\eqref{eq:pwave}, 
we first adopt hard momentum cutoff regularization, 
for which the required regulated value from 
Eq.~\eqref{eq:d0L} 
is
$\d(0) = \frac{\L}{\p}$. 
Renormalization is carried out by comparing 
$f_1(k)$
with the general amplitude in 
Eq.~\eqref{eq:f1PE}, 
and matching to the 
$p$-wave scattering length 
$a_1$
in 
Eq.~\eqref{eq:ERE1}. 
The result is the running coupling 
\begin{equation}
c_2^{(p)}(\L) 
= 
\left( \frac{2 \L}{\p} - \frac{1}{a_1} \right)^{-1}
.\end{equation}
This running is required to maintain the 
$\L$-independence of the scattering length 
$a_1$.

With NDR, 
the renormalization is simpler due to the regulated value 
$\d(0) = 0$
from 
Eq.~\eqref{eq:NDRd0}. 
The 
$c_2^{(p)}$
coefficient is finite in this scheme,
and must have the value 
$c_2^{(p)} = - (a_1)^{-1}$
to match the 
$p$-wave scattering amplitude. 
In NDR, 
we thus have simple scale-independent relations between coefficients of contact operators and the self-adjoint extension parameters
of a parity-even point interaction
\begin{equation}
c_0 
= 
\frac{1-\a}{\d}
\quad \text{and} \quad
c_2^{(p)} 
=
\frac{\d}{1+\a}
\label{eq:PCcs}
.\end{equation}

\subsection{Parity-Odd Contact Interactions}
\label{s:PV}

Next, 
we compute the $T$-matrix for the set of parity-odd contact interactions in 
Table~\ref{t:Ops}. 
We also include the delta-function in the set, 
because
renormalization will generate this interaction even when its bare coupling vanishes. 
The computation is carried out using two different DR schemes, 
and connection is made with some earlier results for parity-violating point interactions.

For the set of parity-odd contact interactions plus the delta-function interaction, 
the momentum-space Schr\"odinger equation~\eqref{eq:pSchro}
produces
\begin{equation}
\phi_\ell(p)
=
\phi_{\ell, \text{inc}}(p)
+ 2 \, 
\frac{c_0 \psi_\ell(0) {-} \mathbb{c}_1 \psi_\ell^{\prime}(0) {+} i p \, \mathbb{c}^*_1 \psi_\ell(0)}{p^2 - k^2 - i \epsilon}
\label{eq:phiPO}
,\end{equation}
which has been compactly written using the complex combination of coefficients
\begin{equation}
\mathbb{c}_1 = c_1 + i \, \widetilde{c}_1
.\end{equation}
Using Eq.~\eqref{eq:phiout}, 
we identify the partial-wave $T$-matrix elements as
\begin{equation}
k \, \mathbb{T}_{0 \, \ell}
= 
c_0 \psi_\ell(0) - \mathbb{c}_1 \psi_\ell^{\prime}(0) 
\quad \text{and} \quad
\mathbb{T}_{1 \, \ell} 
=
i \, \mathbb{c}^*_1 \psi_\ell(0)
.\end{equation}
These involve linear combinations of the wavefunction and its derivative at the origin, 
which themselves are related to the 
$T$-matrix elements by 
Eq.~\eqref{eq:psizeros}.  
Eliminating the wavefunction and its derivative at the origin, 
we obtain the set of two equations 
\begin{align}
k \, \mathbb{T}_{0 \, \ell}
&=
\phantom{i} c_0 \, \big( \d_{0 \, \ell} + i \, \mathbb{T}_{0 \, \ell}  \big)
-
\mathbb{c}_1 \big[ i k \, \d_{1 \, \ell} + 2 i \, \mathbb{T}_{1 \, \ell} \, I_2(k)  \big]
,\notag \\
\mathbb{T}_{1 \, \ell} 
&= 
i \, \mathbb{c}_1^* \left( \d_{0 \, \ell} + i \, \mathbb{T}_{0 \, \ell} \right)
,\end{align}
for the unknown $T$-matrix elements, 
where there is one set for each value of 
$\ell$.
Solving these linear equations, 
we deduce the partial-wave 
$T$-matrix 
\begin{equation}
\mathbb{T}
{=} 
\frac{1}{k {-} i \big[ c_0 {+} 2 | \mathbb{c}_1|^2 \,  I_2(k) \big]}
\begin{pmatrix}
c_0 {+} 2 | \mathbb{c}_1|^2 \,  I_2(k) 
& 
- i k \mathbb{c}_1 
\\
i k \, \mathbb{c}_1^*
& 
 i k | \mathbb{c}_1|^2 
\end{pmatrix}
\label{eq:TPOdd}
.\end{equation}

Before carrying out renormalization in specific regularization schemes,
note that the angle
$\Phi$
is determined by a scheme-independent relation. 
From 
Eq.~\eqref{eq:angles}, 
we have
\begin{equation}
\Phi
=
- \arg \mathbb{c}_1
= 
-
\tan^{-1}
\frac{\widetilde{c}_1}{c_1}
\label{eq:phase}
.\end{equation}
The same is not true of the mixing angle, 
because it satisfies the relation
\begin{equation}
k \cot \Theta
=
\frac{c_0 + 2 | \mathbb{c}_1|^2 \left[ I_2(k) -  \frac{i k}{2}  \right] }{2 |\mathbb{c}_1|}
,\end{equation}
which depends on the scheme used to regulate the integral
$I_2(k)$.

\subsubsection{Renormalization with Dimensional Regularization} %
\label{sec:RDRRR} 

In dimensional regularization with the PDS scheme, 
Eq.~\eqref{eq:I2PDS} shows that 
$I^\text{PDS}_2(k) = \frac{ik}{2} + \m$, 
where 
$\m$
is the dimensional regularization scale.%
\footnote{
As the $T$-matrix in 
Eq.~\eqref{eq:TPOdd}
depends only on 
$I_2(k)$, 
its renormalization using the PDS scheme is consequently the same as with a hard momentum cutoff
$\L$, 
provided we make the substitution 
$\m \to \frac{\L}{\p}$, 
see Eqs.~\eqref{eq:I2L} and \eqref{eq:I2PDS}. 
}
The three running couplings
$c_0(\m)$, 
$c_1(\m)$, 
and
$\widetilde{c}_1(\m)$
of the effective theory require renormalization conditions. 
The $T$-matrix maintains only a single pole, 
and a natural choice is to fix the energy of the pole
$E_0 = - \frac{\k_0^2}{2m}$.  
From 
Eq.~\eqref{eq:TPOdd}, 
this renormalization condition translates into
\begin{equation}
\k_0
= 
\frac{c_0(\m) + 2 \m \,  | \mathbb{c}_1(\m)|^2}{1 + |\mathbb{c}_1(\m)|^2}
\label{eq:PNCcs}
.\end{equation}
Note that 
$\k_0$
need not be positive; 
a bound-state pole and an antibound-state pole are both possible renormalization conditions.
The time-reversal violating phase 
$\Phi$
is a physical parameter that can be deduced from knowledge of the scattering matrix, 
thus we additionally enforce 
Eq.~\eqref{eq:phase}
as a renormalization condition. 
Finally, 
the mixing angle is another physical parameter, 
for which we enforce the condition
\begin{equation}
k \cot \Theta
\equiv
\frac{1}{a_\Theta}
+ 
\cdots
=
\frac{c_0 (\m) + 2 \m \, | \mathbb{c}_1 (\m)|^2}{2 |\mathbb{c}_1(\m)|}
.\end{equation}
Here, 
we have introduced the mixing length 
$a_\Theta$
as an abbreviation, 
but also to emphasize its momentum independence. 
These three conditions enable us to determine the coefficients
\begin{eqnarray}
c_0(\m)
&=&
\k_0
+
\frac{\k_0 - 2 \m}{ \left(  \k_0 \, a_\Theta  \right)^2}
\left[ 1 - \sqrt{1 - \left(  \k_0 \, a_\Theta \right)^2}
\, \right]
^2
,\notag \\
\mathbb{c}_1
&=&
\frac{e^{ - i \Phi}}{ \k_0 \, a_\Theta}
\left[ 1 - \sqrt{1 - \left( \k_0 \, a_\Theta \right)^2}
\, \right]
\label{eq:runs}
,\end{eqnarray}
where only 
$c_0(\m)$
is required to be a running coupling to keep the 
$T$-matrix 
$\m$ independent.%
\footnote{
Note that an additional set of solutions has been ruled out by requiring
$\mathbb{c}_1 \to 0$
in the limit that
$\Theta \to 0 \, \, \text{mod} \, \, \p$, 
or equivalently when
$a_\Theta \to 0$. 
}
Given the matching conditions, 
moreover, 
the condition 
$| \k_0 \, a_\Theta | \leq 1$
is required for a non-perturbative solution.

Explicit computation of the 
the renormalized 
$T$-matrix produces
\begin{equation}
\mathbb{T}
=
\frac{1}{\frac{k}{\k_0} - i}
\begin{pmatrix}
1 
+ 
\frac{i k}{ 2 \k_0}
\c A
& 
- i e^{-i \Phi} \,
\frac{k \, a_\Theta}{2} 
\\
 i e^{ i \Phi} \, \frac{k \, a_\Theta}{2}
& 
\frac{i k}{ 2 \k_0}
\c A
\end{pmatrix}
\label{eq:TRGinv}
,\end{equation}
where
$\c A
=
1 - \sqrt{1 - ( \k_0 \, a_\Theta)^2}
$.
The corresponding eigenstate scattering amplitudes are
\begin{equation}
f_\pm
=
\frac{\k_0
\left[
1 {\pm} \sqrt{1 {+} (k \, a_\Theta)^2} \, 
\right]
{+}
i k
\left[ 
1 {-}  \sqrt{1 {-} ( \k_0 \, a_\Theta)^2} \, 
\right]
}{2 \left( k - i \k_0 \right)}
.\end{equation}
In accordance with the 
$\d = 0$
results of 
Sec.~\ref{sec:onepole}, 
only 
$f_+$
maintains a pole at 
$k = i \k_0$; 
and, 
with 
$a_\Theta \to 0$
specifying the limit of a parity-even interaction, 
the other amplitude vanishes
$f_- \to 0$
in that limit. 
The wavefunction at the pole can be found 
(up to a constant of proportionality) 
from taking the residue of 
$\psi^{(+)}(x)$
at 
$k = i \k_0$, 
which leads to%
\footnote{
Either the right- or left-traveling solution 
$\psi^{(\pm)}(x)$ 
can be used to deduce the wavefunction at the pole. 
That obtained from the left-traveling solution agrees 
with 
Eq.~\eqref{eq:psipole}
up to a complex-valued constant of proportionality. 
Additionally, the wavefunction agrees with the one obtained by directly enforcing 
the joining conditions in 
Eq.~\eqref{eq:SAE1D} 
on a bound-state solution
$\Psi(x)$. 
In terms of the self-adjoint extension parameters, 
one readily finds
$\Psi(x) \propto e^{ - \k_0 |x|} \left[ \theta(-x) \, \a + \theta(x) \,  e^{i \phi}  \right]$, 
and can be rewritten in the form of 
Eq.~\eqref{eq:psipole}
by utilizing the relations in 
Eq.~\eqref{eq:ANGLES0}.
} 
\begin{multline}
\Psi(x)
\propto
e^{ - \k_0 |x|}
\Big[ 
\sqrt{1 - (\k_0 \, a_\Theta)^2} \, + 1 + \k_0 \, a_\Theta \, e^{- i \Phi}
\\
+ \sgn(x) 
\left(
\sqrt{1 - (\k_0 \, a_\Theta)^2} \, - 1 - \k_0 \, a_\Theta \, e^{i \Phi}
\right)
\Big] 
\label{eq:psipole}
,\end{multline}
and is a superposition of parity eigenstates provided 
$a_\Theta \neq 0$.
When 
$\k_0 > 0$, 
the wavefunction can be normalized,
and the pole corresponds to a bound state; 
whereas, 
for 
$\k_0 < 0$, 
the pole corresponds to an antibound state.

Renormalization using the NDR scheme is very similar and reproduces 
$\mathbb{T}$
above. 
From Eq.~\eqref{eq:I2NDR}, 
we note that 
$I^\text{NDR}_2(k) = \frac{ik}{2}$, 
and readily obtain the NDR results from those in the PDS scheme by setting 
$\m = 0$. 
The only modification is thus 
$c_0^\text{NDR} = c_0^\text{PDS}(\m{=}0)$
in Eq.~\eqref{eq:runs}. 
Without scale dependence, 
moreover,
we can directly relate the coefficients of contact operators to the 
self-adjoint extension parameters. 
We have already determined 
$\d = 0$
by comparing with 
Sec.~\ref{sec:onepole}, 
which is further confirmed by noting that
$k \cot \Theta$
is non-zero at threshold.  
The renormalization conditions provide three non-linear equations 
relating the coefficients of contact operators to the three independent self-adjoint extension parameters
$\a$, 
$\be$, 
and
$\phi$. 
In NDR, 
we find
\begin{eqnarray}
c_0 
&=& - 
\frac{2 \be}{\a+\a^{-1} + 2 \cos \phi}
,\notag \\
c_1
&=& \phantom{-}
\frac{\a - \a^{-1}}{\a+\a^{-1} + 2 \cos \phi}
,\notag \\
\widetilde{c}_1
&=& \phantom{-}
\frac{2 \sin \phi}{\a+\a^{-1} + 2 \cos \phi}
\label{eq:PVrelations}
.\end{eqnarray}
From the multiple solutions, 
we choose the set based on the limit of an interaction that is parity even and time-reversal even, 
for which 
$c_0 = - \frac{\be}{2}$.%
\footnote{
Even after this choice, 
there is an overall sign ambiguity so that the related set
$\{ c_0, -c_1, -\widetilde{c}_1 \}$
is also a solution. 
The signs of both parity-odd coefficients can be flipped by redefining the $x$-axis to be reflected about the origin.
}

\subsubsection{Classical Scale Symmetry}%

Now we restrict to the case 
$c_0 = 0$, 
and retain the two parity-odd operators with coefficients
$c_1$
and
$\widetilde{c}_1$
in 
Table~\ref{t:Ops}. 
At the classical level, 
the Hamiltonian has a homogeneous scale transformation. 
Quantum mechanically, 
the higher-dimensional operators will renormalize lower dimensional ones, 
unless protected by a symmetry. 
As the scale symmetry is anomalous at the quantum level, 
the delta-function interaction should be generated in renormalizing the theory, 
even though the bare coupling 
$c_0$
was chosen to vanish. 
This is a way to rephrase the findings of 
Refs.~\cite{SEBA1986,Al-Hashimi:2015nva,Camblong:2019bfr}. 
Indeed starting with such a vanishing bare coupling, 
we have the renormalization condition for the pole location as
\begin{equation}
\k_0 = \frac{ 2 \m | \mathbb{c}_1(\m)|^2}{1 + |\mathbb{c}_1(\m)|^2}
,\end{equation}
in the PDS scheme.%
\footnote{
With 
$c_0 = 0$
in the NDR scheme, 
classical scale symmetry is preserved. 
Consequently, the scattering matrix does not possess a pole and the scattering amplitudes are anomalous, 
see Sec.~\ref{s:SAEscale}. 
It is a tacit assumption that such behavior is excluded by the renormalization conditions. 
When 
$c_0 \neq 0$, 
by contrast, 
there is no scale symmetry 
and NDR provides a perfectly reasonable regularization scheme. 
} 
This form excludes the possibility of an antibound-state pole, 
because only 
$\k_0 > 0$
is permitted. 
The relative phase is scheme independent as before, 
which translates to 
$\mathbb{c}_1(\m) = e^{- i \Phi} | \mathbb{c}_1(\m)|$. 
The running of the modulus coupling is given by
\begin{equation}
|\mathbb{c}_1(\m)|
=
\sqrt{\frac{\k_0}{2 \m - \k_0}}
\label{eq:c1run}
.\end{equation}
This running coupling then renders the mixing angle in the form
\begin{equation}
k \cot \Theta
=
\m \, \sqrt{\frac{\k_0}{2 \m - \k_0}}
\label{eq:mixrun}
,\end{equation}
which is, 
however,
scale dependent. 
Consequently, 
the $T$-matrix for the theory with 
$c_1(\m)$
and
$\widetilde{c}_1(\m)$
cannot be renormalized.%
\footnote{
The same conclusion can be reached by alternately 
enforcing the renormalization condition for the mixing angle first. 
In this case, 
we arrive at a running coupling
$|\mathbb{c}_1(\m)| \propto \m^{-1}$. 
With this coupling, 
the pole of the 
$T$-matrix remains 
$\m$ dependent
and the scattering matrix also cannot be renormalized. 
} 
An additional interaction is required with a running coupling that 
can render the 
scattering matrix to be scale independent. 

Curiously,
Ref.~\cite{Camblong:2019bfr}
does not obtain a scattering matrix that is invariant under renormalization group evolution;
instead, 
the limit
$\m \to \infty$
is taken.%
\footnote{
In an even number of dimensions, 
residual logarithmic dependence on the renormalization scale would obviously preclude taking this limit.
While there are no logarithms in odd dimensions, 
renormalization scale dependence still signals the need for an additional operator to renormalize 
the scattering matrix. 
The running couplings in 
Eq.~\eqref{eq:runs} 
exhibit that the coefficient of the delta-function interaction is relevant, 
while those of the first-derivative interactions are marginal. 
Without logarithmic running, 
the marginal operators have a finite, 
scale-independent renormalization. 
} 
With the running coupling in 
Eq.~\eqref{eq:c1run}, 
the bound-state pole  
remains fixed in this limit by design. 
The mixing angle in 
Eq.~\eqref{eq:mixrun}, 
however, 
has the limit
$\Theta \left(\m{\to}\infty \right) = 0$, 
for which parity invariance emerges. 
The corresponding limit of the partial-wave
$T$-matrix
\begin{equation}
\mathbb{T} \left(\m {\to} \infty \right)
=
\frac{1}{\frac{k}{\k_0} - i \,}
\begin{pmatrix}
1 & 0 \\ 0 & 0 
\end{pmatrix}
,\end{equation} 
is that of an energy-independent 
$s$-wave contact interaction. 
The correct conclusion is that the parity-violating operators evolve to a parity conserving 
$\d(x)$ interaction
as 
$\m \to \infty$. 
This is the missing interaction needed to renormalize the theory, 
which is clear from renormalization theory: 
the higher-dimensional operators renormalize lower-dimensional ones. 
Once we accept that the scale symmetry is anomalous, 
a delta-function interaction cannot be omitted in the renormalization
of the scattering matrix. 
In Sec.~\ref{sec:RDRRR}, 
a scale-independent scattering matrix 
Eq.~\eqref{eq:TRGinv} 
is obtained with inclusion of the 
delta-function interaction; 
and, parity sensibly remains broken.

\subsection{Contact Interactions up to Second-Derivative Order}
\label{s:aciutsdo}

Lastly, 
the scattering solution is obtained using the set of four energy-independent contact interactions shown in 
Table~\ref{t:Ops}. 
While the calculation can be performed in any regularization scheme, 
the expressions are rather cumbersome to display in all but the NDR scheme. 
This scheme automatically subtracts power-law divergences, 
for which the coefficients of contact operators are finite.

The momentum-space Schr\"odinger equation%
~\eqref{eq:pSchro}
now includes all terms in 
Eqs.~\eqref{eq:phiPE}
and
\eqref{eq:phiPO}. 
The partial-wave $T$-matrix elements are thus given by
\begin{eqnarray}
k \, \mathbb{T}_{0 \, \ell} 
&=&
c_0  \psi_\ell(0) 
- 
\mathbb{c}_1 \psi_\ell^{\prime}(0) 
,\notag \\
\mathbb{T}_{1 \, \ell}
&=&
i \, \mathbb{c}^*_1 \psi_\ell(0)  - i \, c_2^{(p)} \psi_\ell^\prime(0) 
.\end{eqnarray}
Eliminating the origin values of the wavefunction and its derivative using 
Eq.~\eqref{eq:psizeros}, 
we obtain the equations
\begin{align}
k \, \mathbb{T}_{0 \, \ell} 
&=
c_0 ( \d_{0 \, \ell} + i \, \mathbb{T}_{0 \, \ell}  )
- 
\mathbb{c}_1 ( i k \, \d_{1 \, \ell} - k \, \mathbb{T}_{1 \, \ell} )
,\notag \\
\mathbb{T}_{1 \, \ell}
&=
i \, \mathbb{c}^*_1 \big( \d_{0 \, \ell} + i \, \mathbb{T}_{0 \, \ell}  \big)  
- 
i \, c_2^{(p)} ( i k \, \d_{1 \, \ell} - k \, \mathbb{T}_{1 \, \ell} )
\label{eq:T6}
,\end{align} 
where we have used Eq.~\eqref{eq:I2NDR} for the momentum integral in NDR.
Solutions of Eq.~\eqref{eq:T6} for the $T$-matrix elements can be written as
\begin{align}
\mathbb{T}_{0 \, \ell} 
&=
\frac{
\left[
\big( 1 - i k \, c_2^{(p)}  \big)
c_0
+ i k | \mathbb{c}_1|^2
\right] \d_{0 \, \ell} 
- 
i k \, \mathbb{c}_1  \d_{1 \, \ell} 
}
{\big( 1 - i k \, c_2^{(p)}  \big) \left( k - i  c_0 \right) + k | \mathbb{c}_1|^2 }
,\notag \\
\mathbb{T}_{1 \, \ell}
&=
\frac{
i k \, \mathbb{c}^*_1 \d_{0 \, \ell} 
+ 
\left[
k \, c_2^{(p)} \left( k - i  c_0 \right) + i k | \mathbb{c}_1|^2
\right]
\d_{1 \, \ell} 
}
{\big( 1 - i k \, c_2^{(p)}  \big) \left( k - i  c_0  \right) + k | \mathbb{c}_1|^2 }
.\notag\\
\label{eq:BigT}
\end{align}

As the breaking of parity and time-reversal is solely due to 
$\mathbb{c}_1$, 
it is not surprising that the relative phase is determined by the same condition as in 
Sec.~\ref{sec:RDRRR}. 
Thus, 
we have 
\begin{equation}
\mathbb{c}_1 = e^{- i \Phi} | \mathbb{c}_1|
\quad
\text{and}
\quad
\frac{\widetilde{c}_1}{c_1} = \frac{2 \sin \phi}{\a - \g}
\label{eq:phase1}
,\end{equation}
where the former equation expresses the phase of 
$\mathbb{c}_1$
in terms of the physically measurable parameter 
$\Phi$, 
and the latter equation gives the relation between the coefficients of contact operators 
and the self-adjoint extension parameters of the general point interaction. 
Going further, 
the mixing angle can be determined
from 
Eq.~\eqref{eq:angles}, 
from which we find
\begin{equation}
k \cot \Theta
=
\frac{c_0 -  k^2 \, c_2^{(p)}}
{2 |\mathbb{c}_1|}
\label{eq:angle1}
,\end{equation}
and is appropriately a linear function of the energy. 
The locations of the poles of the scattering matrix provide the final renormalization conditions. 
From 
Eq.~\eqref{eq:BigT}, 
the common denominator of all $T$-matrix elements
is proportional to 
$\left[k - i \k_+ \right] \, \left[k - i k_- \right]$, 
where
\begin{multline}
\k_\pm
=
-\frac{1 + | \mathbb{c}_1|^2 - c_2^{(p)} c_0}{2 \, c_2^{(p)}}
\\
\pm 
\frac{
\sqrt{\left[ 1 + | \mathbb{c}_1|^2 - c_2^{(p)} c_0 \right] ^2 + 4 c_2^{(p)} c_0}}{2 \, c_2^{(p)}}
\label{eq:6cs}
.\end{multline}
Thus,
all $T$-matrix elements have two poles, 
and the associated bound or antibound states are of indefinite parity. 
In turn, 
Eqs.~\eqref{eq:roots} and~\eqref{eq:ANGLES}
allow us to deduce the coefficients of contact operators in terms of the self-adjoint extension parameters. 
Straightforward algebra reveals the sought-after relations%
\footnote{
As with 
Eq.~\eqref{eq:PVrelations} 
above, 
there is a second set of solutions differing only by   
$c_1$
and 
$\widetilde{c}_1$
both multiplied by 
$-1$. 
}
\begin{align}
c_0
&=
- \frac{2 \be}{\a + \g  + 2 \cos \phi}
,\quad 
c_2^{(p)}
= 
\frac{2 \d }{\a + \g  + 2 \cos \phi}
,\notag \\
c_1
&=
\phantom{-} \frac{\a - \g}{\a + \g + 2 \cos \phi}
,\quad
\widetilde{c}_1
=
\frac{2 \sin \phi}{\a + \g + 2 \cos \phi}
\label{eq:dictionary1}
.\end{align}

\section{Harmonic Trap with a Point Interaction in One Dimension} %
\label{s:SHO1D}

Having treated one-dimensional scattering from a general point interaction in the absence of a long-range potential, 
we add a harmonic oscillator potential. 
This is a straightforward application of the methods from Sec.~\ref{s:SHO3} and Sec.~\ref{s:SAE1D}. 
The Hamiltonian is taken to be
\begin{equation}
H = \frac{p^2}{2m} + \frac{1}{2} m \o^2 x^2
,\end{equation}
on the punctured line
$x \in \mathbb{R}/\{0\}$.
The self-adjoint extension to 
$x = 0$
is made through the joining conditions in 
Eq.~\eqref{eq:SAE1D}. 
We assume the separation of long- and short-range interactions,
so that the oscillator potential has no effect on the physics producing the point interaction, 
see Fig.~\ref{f:SHO} for the analogous situation on the half line. 
As a result, 
the self-adjoint extension parameters take on their values for 
$\o = 0$. 
These values give a complete characterization of the one-dimensional scattering problem
at low energies.
The scattering lengths, 
mixing pattern, 
and time-reversal violating phase are encoded in the extension parameters.

On the punctured line, 
the normalizable solutions to the Schr\"odinger eigenvalue problem 
$H \psi_E(x) = E \, \psi_E(x)$
must be of the form
\begin{multline}
\psi_E(x)
= 
\theta(-x)
\, N_- \, U \big( {-}\tfrac{E}{\o}, {-} \sqrt{2 m \o} \, x \big)
\\
+ 
\theta(x)
\, N_+ \,  U \big( {-}\tfrac{E}{\o}, \sqrt{2 m \o} \, x \big)
\label{eq:SHO1d}
,\end{multline}
where  
$U(a,z)$
is the parabolic cylinder function employed in 
Sec.~\ref{s:SHO3}. 
The amplitudes on either side of the origin are
$N_\pm$, 
which are generally complex numbers. 
The solution introduces four real parameters, however, 
the normalization condition on the wavefunction 
and the overall phase convention reduce the number of free parameters to two. 
The amplitudes must be determined by enforcing the joining condition in 
Eq.~\eqref{eq:SAE1D}. 
As there are four self-adjoint extension parameters in the general point interaction, 
we can anticipate that the solution is over determined, 
leading to a quantization condition on the energy 
$E$.

For the oscillator solution defined piecewise in 
Eq.~\eqref{eq:SHO1d}, 
the joining conditions translate into the pair of equations
\begin{multline}
N_+ 
\begin{pmatrix}
\sqrt{2 m\o\,}  \,
U'\left(-\frac{E}{\o} , 0\right)
\\
U\left(-\frac{E}{\o} , 0\right)
\end{pmatrix}
=
\\
e^{ i \phi} 
N_-
\begin{pmatrix}
-\a \sqrt{2 m\o\,}  \,
U'\left(-\frac{E}{\o} , 0\right)
+ 
\be \, U\left(-\frac{E}{\o} , 0\right)
\\
-\d \sqrt{2 m\o\,}  \,
U'\left(-\frac{E}{\o} , 0\right)
+ 
\g \,U\left(-\frac{E}{\o} , 0\right)
\end{pmatrix}
.\end{multline}
From these equations and the function values in Eq.~\eqref{eq:UU'}, 
one finds the spectrum condition
\begin{equation}
\a + \g
=
-
\frac{\be}{2 \sqrt{m \o\,}}
\
\frac{\Gamma\left(\frac{1}{4}{-}\frac{E}{2\o}\right)}{\Gamma\left(\frac{3}{4} {-}\frac{E}{2\o} \right)}
-  
2 \sqrt{m\o\,} \, \d
\
\frac{\Gamma\left(\frac{3}{4} {-}\frac{E}{2\o} \right)}{\Gamma\left(\frac{1}{4}{-}\frac{E}{2\o}\right)}
\label{eq:SHO1}
.\end{equation}
Solutions of this transcendental equation determine the energy 
$E$. 
Notice that such solutions are independent of the time-reversal violating parameter 
$\phi$,  
which therefore only enters the wavefunction.

As written above, 
the spectrum condition depends on three independent self-adjoint extension parameters, 
which can be chosen as
$\a$, 
$\g$, 
and 
$\d$, 
using the relation in Eq.~\eqref{eq:det}.
In turn, 
these parameters can be expressed in terms of quantities accessible from the scattering matrix.
Using the latter relations, 
the equation for the spectrum becomes
\begin{multline}
\left[ \Gamma\left(\tfrac{3}{4} -\tfrac{E}{2\o} \right) 
- 
\frac{\k_+}{2 \sqrt{m \o\,}} \Gamma\left(\tfrac{1}{4}-\tfrac{E}{2\o}\right) \right]
\\
\times
\left[ \Gamma\left(\tfrac{3}{4} -\tfrac{E}{2\o} \right) 
- 
\frac{\k_-}{2 \sqrt{m \o\,}} \Gamma\left(\tfrac{1}{4}-\tfrac{E}{2\o}\right) \right]
=0
,\end{multline}
which is independent of the pattern of mixing between partial waves and the relative phase between them. 
Only the two poles of the 
$S$-matrix,  
$\k_\pm$
given in 
Eq.~\eqref{eq:roots},
are required.

To further exhibit the solution, 
it is useful to investigate limiting cases. 
For the parity-even point interaction detailed in
Sec.~\ref{s:PE}, 
each pole of the $S$-matrix is determined by a partial-wave scattering length. 
In terms of scattering lengths, 
the spectrum condition for the parity-even point interaction is satisfied when either
\begin{equation}
2 \sqrt{m \o\,} \
\frac{\Gamma\left(\frac{3}{4} -\frac{E}{2\o} \right)}{\Gamma\left(\frac{1}{4}-\frac{E}{2\o}\right)}
= 
\frac{1}{a_0},
\, \text{ or } \,
\frac{1}{a_1}
\label{eq:SHO1s} 
.\end{equation}
These formulas are analogous to the three-dimensional case
Eq.~\eqref{eq:TRAP}.
The three-dimensional case, 
however, 
has only
$s$-wave interactions between particles;
whereas, 
the one-dimensional case above includes both 
$s$- and $p$-wave interactions in uncoupled channels.%
\footnote{
For the delta-function interaction 
$V = - \frac{c_0}{m} \, \d(x)$, 
the self-adjoint extension parameters are given by 
$\a = \g = 1$, 
$\d = \phi = 0$, 
and
$\be = - 2 c_0$, 
with the scattering length
$a_0 = (c_0)^{-1}$. 
The spectrum from 
Eq.~\eqref{eq:SHO1}
is then determined by
$c_0 
= 
2 \sqrt{m \o \,} \ \frac{\Gamma\left(\frac{3}{4}-\frac{E}{2\o}\right)}{\Gamma\left(\frac{1}{4} -\frac{E}{2\o} \right)}
$, 
which is the one-dimensional formula derived in 
Ref.~\cite{Busch:1998}.
}

As a curiosity,  
the final limit we take is that of the scale-invariant point interaction 
Sec.~\ref{s:SAEscale}
in the harmonic oscillator potential.%
\footnote{
While the harmonic oscillator introduces a scale, 
there is a dynamical 
$SO(2,1)$
symmetry%
~\cite{PhysRevA.55.R853}. 
The spectrum-generating algebra, 
for example, 
leads to evenly spaced levels with 
$\D E = 2 \o$.
This is essentially a classical scale symmetry that is broken by quantum effects, 
and our results indirectly confirm the anomaly. 
} 
Accordingly, 
we restrict the self-adjoint extension parameter
$\g = \a^{-1}$,
and attempt to set 
$\be = \d = 0$.
From 
Eq.~\eqref{eq:SHO1}, 
however,
the equation determining the spectrum 
would seem to produce a contradiction 
\begin{equation}
\frac{\a + \a^{-1}}
{\Gamma\left(\frac{1}{4}{-}\frac{E}{2\o}\right) \Gamma\left(\frac{3}{4} {-}\frac{E}{2\o} \right)}
=
-
\frac{\frac{\be}{2 \sqrt{m \o\,}}}{\Gamma\left(\frac{3}{4} {-}\frac{E}{2\o} \right)^2}
{-}  
\frac{2 \sqrt{m\o\,} \, \d}{\Gamma\left(\frac{1}{4}{-}\frac{E}{2\o}\right)^2}
,\end{equation}
because the right-hand side vanishes, 
whereas 
$\a + \a^{-1}$
cannot vanish. 
The spectrum condition can only be satisfied at particular energies 
$E$
at which the denominator has a pole. 
The gamma-function duplication formula%
~\cite[Eq.~(5.5.5)]{NIST:DLMF}
\begin{equation}
\Gamma(z) \Gamma \big(\tfrac{1}{2} + z\big) 
= 
\sqrt{\p}
\, 2^{1-2z} \,
\Gamma(2z)
,\end{equation}
allows us to focus on a single gamma function rather than the product of two. 
With 
$\be = \d = 0$, 
the equation for the spectrum becomes 
\begin{equation}
\frac{\a + \a^{-1}}{\Gamma\big(\frac{1}{2} - \frac{E}{\o}\big)}
=
0
,\end{equation} 
and can thus be satisfied when 
$\frac{1}{2} - \frac{E}{\o} = - n$ 
for 
$n \in \mathbb{Z}^+$. 
Of course,
the condition  
$E = \o (n + \frac{1}{2})$
yields the spectrum of the harmonic oscillator without any point interaction. 
Consistency of the quantum mechanical eigenvalue equation thus implies that the only 
scale-invariant point interaction must be trivial.

\section{Summary}
\label{s:summy}

We explore connections between low-energy scattering, 
self-adjoint extensions, 
and contact interactions. 
To conclude, 
we summarize the major results and indicate a few remaining questions. 
The case of short-range $s$-wave interactions is reviewed in terms of the 
impenetrable self-adjoint extension of the reduced-radial Hamiltonian in 
Sec.~\ref{sec:swave}. 
The self-adjoint extension parameter accounts for the scattering length. 
As a simple application, 
we extend the analysis to harmonically confined particles. 
These introductory considerations pave the way to consider 
self-adjoint extensions on the punctured line in 
Sec.~\ref{sec:point}.
The general
$S$-matrix for a particle scattering from finite-range potential is detailed in 
Eq.~\eqref{eq:T}. 
The result is written in terms of eigenstate scattering amplitudes
$f_\pm$, 
a mixing angle
$\Theta$,
and a relative phase
$\Phi$. 
These quantities are related to the self-adjoint extension parameters 
of the one-dimensional point interaction through 
Eqs.~\eqref{eq:EVPhases} and \eqref{eq:ANGLES}. 
In particular, 
the quantity 
$k \cot \Theta$
is found to be a linear function of the energy. 
The formulation of one-dimensional scattering in terms of waves of definite parity is elaborated on 
in~\ref{s:parity}.

Short-range interactions can be described by a potential consisting 
of a tower of contact interactions, 
provided the calculations are regulated and renormalized.
In one dimension, 
all such contact interactions up to second-derivative order are given in 
Table~\ref{t:Ops}.
The partial-wave basis is utilized for these operators, 
and their properties are shown. 
The Schr\"odinger equation is solved for sets of energy-independent contact interactions
by generalizing the momentum-space method of 
Ref.~\cite{Jackiw:1991je}. 
The hard momentum cutoff regularization and dimensional regularization schemes 
that we employ are detailed in~\ref{s:DimReg}. 
As an example, 
all four energy-independent contact interactions are used to determine the 
$T$-matrix using the NDR scheme in 
Sec.~\ref{s:aciutsdo}. 
The physical parameters of the scattering matrix are related to the coefficients of contact operators in Eqs.~\eqref{eq:phase1}--\eqref{eq:6cs}. 
Additionally, 
we provide the dictionary that translates between the self-adjoint extension parameters and 
coefficients of contact operators in 
Eq.~\eqref{eq:dictionary1}. 
When including higher-dimensional 
(marginal and irrelevant)
contact operators, 
we emphasize that suitable renormalization conditions must be enforced to arrive at a
scale-independent scattering matrix. 
In particular,  
the single-derivative operators of 
Sec.~\ref{s:PV}
introduce mixing of parity that 
has not been previously addressed.

There are a few points of interest for further work. 
With only two channels, 
the pattern of mixing in one-dimensional scattering is considerably simpler than in two and three dimensions. 
This simplicity suggests that there may be a form of Levinson's theorem for the eigenstate phase shifts of parity non-symmetric interactions. 
In another direction, 
self-adjoint extensions on the punctured line with maximal time-reversal violation
lead to the very curious behavior of  
$S$-matrix elements in
Eq.~\eqref{eq:Swhat}. 
In this limit, 
the theory can support a bound state, 
for example, 
without the 
$S$-matrix exhibiting a pole. 
It is not clear whether a low-energy theory of contact interactions can reproduce such behavior, 
and further work exploring the consequences of time-reversal violation seems warranted. 
Lastly, 
our analysis is restricted to those contact operators that are additionally self-adjoint, 
which excludes those with energy dependence. 
The description of low-energy scattering in effective theories can be improved, 
however,
by including energy-dependent contact interactions,
such as the interaction that generates an effective range. 
In strict terms,  
a positive effective range cannot be treated non-perturbatively%
~\cite{Beane:1997pk,vanKolck:1998bw}. 
For contact interactions, 
this is a reflection of Wigner's causality bound%
~\cite{Wigner:1955zz,Phillips:1996ae,Phillips:1997xu,Hammer:2009zh,Hammer:2010fw,Beck:2019abp}.  
The topic of energy-dependent point interactions has received renewed attention 
\cite{10.1063/1.5048692,Guliyev2019b,Granet:2022zpc}, 
and there might be insight gained from comparing the perspectives of effective field theory 
and mathematical physics.

Nevertheless, 
we present a comprehensive modern perspective on
quantum mechanics with short-range interactions formulated in one dimension. 
We show that contact interactions can provide a complementary 
description to self-adjoint extensions, 
provided the former are suitably regulated and renormalized. 
We hope that these topics become a natural supplement to introductory 
quantum mechanics courses, 
in preparation for more advanced treatment using quantum field theory.

\section*{CRediT Authorship Contribution Statement}

{\bf Daniel R.~DeSena}: Conceptualization, Formal analysis, Validation.
{\bf Brian C.~Tiburzi}: Conceptualization, Formal analysis, Validation, Writing - Original Draft, Writing - Review \& Editing.

\section*{Declaration of Competing Interest}

The authors declare that they have no known competing financial interests or personal relationships that could have appeared to influence the work reported in this paper.

\section*{Acknowledgements}

This research did not receive any specific grant from funding agencies in the public, commercial, or not-for-profit sectors.

\appendix

\section{Scattering in One Dimension}
\label{s:parity}

Scattering theory in one dimension is reviewed. 
The $S$-matrix is first formulated in the basis of right- and left-traveling waves, 
then converted to the partial-wave basis. 
The case of a parity-even interaction is detailed.

\subsection{S-Matrix in One Dimension}
\label{s:smatrix}

The 
$S$-matrix
provides the most economical description of scattering. 
In one dimension, 
asymptotic incoming and outgoing waves are superpositions of right- and left-traveling waves
\begin{eqnarray}
\Psi_\text{in}(x) 
&\overset{k|x| \gg 1}{=}&
\theta({-}x) \,
\Psi_\text{in}^{(+)} \, e^{+i k x}
{+} 
\theta(x) \, 
\Psi_\text{in}^{(-)} \, e^{- i k x} 
,\notag \\
\Psi_\text{out}(x) 
&\overset{k|x| \gg 1}{=}& 
\theta({-}x) \, 
\Psi_\text{out}^{(-)} \, e^{- i k x} 
{+} 
\theta(x) \,
\Psi_\text{out}^{(+)} \, e^{+i k x}
.\notag\\
\label{eq:inout}
\end{eqnarray}
The outgoing amplitudes are related to the incoming amplitudes by operation with the $S$-matrix%
\footnote{
Note that the right-traveling (left-traveling) wave is thus incident from the left (right).
We label these waves by their propagation direction not location, 
which is then similarly done for the outgoing waves. 
Tracking instead by location,
one would be inclined to write the transformation 
(in our notation)
as
$\begin{pmatrix}
\Psi_\text{out}^{(-)} \\
\Psi_\text{out}^{(+)} 
\end{pmatrix}
=
\S_1  S
\begin{pmatrix}
\Psi_\text{in}^{(+)} \\
\Psi_\text{in}^{(-)} 
\end{pmatrix}
$,
as suggested in 
Refs.~\cite{merzbacher1998quantum,griffiths_introduction_2018}.
While the resulting matrix 
$\widetilde{S} = \S_1 S$
is unitary, 
it is neither unitarily equivalent to 
$S$, 
nor does it reduce to the identity matrix for a free particle. 
}
\begin{equation}
\begin{pmatrix}
\Psi_\text{out}^{(+)} \\
\Psi_\text{out}^{(-)} 
\end{pmatrix}
=
S
\begin{pmatrix}
\Psi_\text{in}^{(+)} \\
\Psi_\text{in}^{(-)} 
\end{pmatrix}
\label{eq:S}
.\end{equation}
The energy of these asymptotic free-particle solutions is defined to be 
$E = \frac{k^2}{2m}$,
with 
$k > 0$
as the scattering momentum. 
This follows from the assumption that the potential is of finite range.

For a parity-even interaction, 
an incoming wave that is reflected about the origin
$\Psi_\text{in}(-x)$
must have the corresponding outgoing solution 
$\Psi_\text{out}(-x)$
obtained from 
Eq.~\eqref{eq:inout}. 
This will be the case provided the
$S$-matrix has the property 
$S = \S_1 S \, \S_1$, 
where
$\S_1$
is defined in Eq.~\eqref{eq:Smatrices}. 
On the other hand, 
imposing time-reversal invariance requires the relation
$\big[ \Psi_\text{in}(x) \big]^* = \Psi_\text{out}(x)$
between incoming and outgoing waves. 
The 
$S$-matrix 
must then obey the relation
$S^T = \S_1 S \, \S_1$, 
where we have used the unitarity condition to identify
$(S^*)^{-1} = S^T$.

Due to linearity, 
the matrix elements of 
$S$
can be obtained by considering incoming right- and left-traveling waves separately. 
For incoming waves with unit amplitude, 
the asymptotic scattering solutions are 
\begin{multline}
\psi^{(\pm)}(x) 
\overset{k|x| \gg 1}{=}
\theta({\mp}x) \big[ e^{ \pm i k x} + R^{(\pm)} \, e^{ \mp i k x} \big] 
\\
+ \theta({\pm}x) \, T^{(\pm)} e^{\pm i k x}
\label{eq:Rinc}
,\end{multline}
which depend on amplitudes
$R^{(\pm)}$
for reflection, 
and  
$T^{(\pm)}$
for transmission. 
With these parameterizations of the solutions for incoming right- and left-traveling waves, 
the form of the
$S$-matrix is
\begin{equation}
S
= 
\begin{pmatrix}
T^{(+)} & R^{(-)}
\\
R^{(+)} & T^{(-)}
\end{pmatrix}
\label{eq:SM}
.\end{equation}
Unitarity of the 
$S$-matrix hinges on probability conservation and orthogonality. 
The conditions
$|R^{(\pm)}|^2 + |T^{(\pm)}|^2 = 1$ 
produce unity along the diagonal of 
$S^\dagger S$. 
Its off-diagonal elements vanish due to orthogonality of the asymptotic right- and left-traveling scattering states 
$\left( R^{(-)} \right)^* T^{(+)}  + \left( T^{(-)} \right)^*  R^{(+)}  = 0$. 
For a parity-even interaction, 
the condition 
$S = \S_1 S\, \S_1$
translates into the requirements 
$R^{(+)} = R^{(-)}$
and
$T^{(+)} = T^{(-)}$. 
With a time-reversal even interaction, 
the condition 
$S^T = \S_1 S \, \S_1$
only imposes the requirement that 
$T^{(+)} = T^{(-)}$.

\subsection{Partial Waves in One Dimension}
\label{s:pwa1d}

Eigenvalues of the $S$-matrix encode the scattering phase shifts.  
To compare with scattering theory, 
we need a one-dimensional analogue of the partial-wave expansion. 
This naturally takes on a rather simple form,%
\footnote{
Despite the simplicity, 
there can be subtleties in practice, 
such as an unusual form for Levinson's theorem and the related Beth-Uhlenbeck formula for the 
second virial coefficient%
~\cite{Camblong:2019obf}.
} 
which has been given in  
Ref.~\cite{10.1119/1.1970982}. 
The analogues of partial waves in one dimension are symmetric and antisymmetric waves, 
which arise from the basic fact that every function can be written as a sum of symmetric and antisymmetric combinations. 
The general point interaction, 
however, 
is not parity invariant. 
As a result, 
the symmetric and antisymmetric waves are coupled%
~\cite{10.1063/1.530481,10.1119/1.18123}. 
To expose the coupling of partial waves, 
we transform the 
$S$-matrix from the basis of 
right- and left-traveling waves 
to that of symmetric and antisymmetric waves. 
Noting, 
for example, 
that the outgoing wave can be written in terms of symmetric and antisymmetric combinations in the form
\begin{multline}
\Psi_\text{out}(x)
=
\frac{1}{\sqrt{2}}
\Bigg[ 
\frac{\Psi^{(+)}_\text{out} + \Psi^{(-)}_\text{out}}{\sqrt{2}} 
\\
+
\sgn(x)  
\frac{\Psi^{(+)}_\text{out} - \Psi^{(-)}_\text{out}}{\sqrt{2}}
\Bigg]
e^{- i k |x|}
\label{eq:PsiOut}
\end{multline} 
we infer a transformation to the partial-wave basis
\begin{equation}
\begin{pmatrix}
\Psi^{(s)}_\text{out} \\
\Psi^{(p)}_\text{out}
\end{pmatrix}
=
U
\begin{pmatrix}
\Psi^{(+)}_\text{out} \\
\Psi^{(-)}_\text{out}
\end{pmatrix}
,\end{equation}
where $(s)$ denotes the symmetric and $(p)$ the antisymmetric wave amplitudes. 
The unitary matrix that carries out the transformation is 
$U = 
\frac{1}{\sqrt{2}}
\begin{pmatrix}
1 & \phantom{-} 1
\\
1 &  - 1
\end{pmatrix}
$.
In the partial-wave basis, 
the 
$S$-matrix from 
Eq.~\eqref{eq:SM}
transforms into 
\begin{equation}
\mathbb{S} 
\equiv
U \, S \, U^\dagger
=
\begin{pmatrix}
\ol T + \ol R 
& 
\D T + \D R 
\\
\D T - \D R
& 
\ol T - \ol R
\end{pmatrix}
\label{eq:SS}
,\end{equation}
where barred quantities are defined to be averages
\begin{eqnarray}
\overline{\c O} 
&=&
\frac{1}{2} \left( \c O^{(+)} + \c O^{(-)} \right)
,\text{ and}
\notag\\
\D \c O 
&=& 
\tfrac{1}{2} \left(  \c O^{(+)} - \c O^{(-)} \right)
\label{eq:avediff}
,\end{eqnarray}
are half of the differences.

For a parity-even interaction, 
we have 
$\D R = \D T = 0$. 
Consequently the 
$S$-matrix 
Eq.~\eqref{eq:SS}
becomes diagonal in the partial-wave basis. 
With an interaction that is time-reversal invariant,  
$\D T = 0$
and
$\mathbb{S}$
becomes an antisymmetric matrix. 
It can be diagonalized in terms of a mixing angle between symmetric and antisymmetric waves%
~\cite{10.1119/1.18123}. 
With breaking of both parity and time reversal,  
the situation becomes more intricate. 
As 
$\mathbb{S} \in U(2)$, 
the partial-wave mixing can generally be diagonalized with an 
$SU(2)$ 
transformation.

In the main text, 
we work in the partial-wave basis. 
The solutions for incoming right- and left-traveling waves 
Eq.~\eqref{eq:Rinc}
can be decomposed into symmetric and antisymmetric outgoing waves
\begin{equation}
\psi^{(\pm)}(x)
\overset{k |x| \gg 1}{=}
e^{\pm i k x}
+
i \, e^{i k |x|}
\left[ 
f^{(\pm)}_0 
\pm  
\sgn(x) \, f^{(\pm)}_1
\right]
\label{eq:f0f1}
.\end{equation}
The symmetric wave is analogous to an 
$s$-wave, 
while the antisymmetric wave is analogous to a 
$p$-wave. 
Note that the first Legendre polynomial is 
$P_1(\cos \theta) = \cos \theta$, 
while the scattering angle is restricted to satisfy 
$\cos \theta = \pm 1$ 
and accordingly obeys the formula 
$\cos \theta = \sgn(x)$. 
The reflected and transmitted amplitudes in Eq.~\eqref{eq:Rinc}
are related to the partial-wave amplitudes
$f_\ell^{(\pm)}$
through the relations
\begin{eqnarray}
R^{(\pm)} 
&=&
i \left( f_0^{(\pm)} - f_1^{(\pm)} \right)
,\notag\\
T^{(\pm)}
&=&
1 + i \left( f_0^{(\pm)} + f_1^{(\pm)} \right)
\label{eq:RTfs}
.\end{eqnarray}
The latter are convenient for expressing the scattering
$T$-matrix, 
which is defined as
\begin{equation}
\mathbb{T}
= 
\frac{\mathbb{S} - \mathbb{1}}{2i}
\label{eq:Tmatrix}
.\end{equation}
In the partial-wave basis,
it has the explicit form
\begin{equation}
\mathbb{T}
=
\begin{pmatrix}
\ol f_0 & \D f_0
\\
\D f_1 & \ol f_1
\end{pmatrix}
\label{eq:Tells}
,\end{equation}
where the barred and difference quantities are defined as in 
Eq.~\eqref{eq:avediff}. 
To fully utilize the partial-wave basis, 
one should employ incoming waves that are symmetric and antisymmetric 
\begin{equation}
\psi_\ell(x)
\equiv 
\frac{\psi^{(+)}(x) +  (-1)^\ell \, \psi^{(-)}(x)}{2} 
,\end{equation}
where 
$\ell = 0$ 
or 
$1$. 
From Eq.~\eqref{eq:f0f1}, 
these scattering solutions have the asymptotic 
$k |x| \gg 1$
behavior 
\begin{equation}
\psi_\ell(x) 
=
i^\ell 
\cos \left( kx - \tfrac{\ell \p}{2} \right)
{+} 
i \, e^{ i k |x| } 
\Big[ \mathbb{T}_{0 \, \ell} {+} \sgn(x) \, \mathbb{T}_{1 \, \ell} \Big]
\label{eq:partialpartial}
,\end{equation}
and have outgoing amplitudes that are
$T$-matrix elements
in the partial-wave basis
Eq.~\eqref{eq:Tells}.

\subsection{Parity-Even Interaction}
\label{s:psi}

For a parity-even interaction, 
we can connect with the partial-wave analysis of
Ref.~\cite{10.1119/1.1970982}. 
In this case, 
the asymptotic scattering solutions must be related by a parity transformation 
$\psi^{(+)}(-x) = \psi^{(-)}(x)$, 
which leads to the requirement
$f_\ell^{(+)} = f_\ell^{(-)}$, 
and consequently
\begin{equation}
\mathbb{T} = \text{diag} \left( f_0, f_1 \right)
\label{eq:Tpwa}
.\end{equation} 
With parity symmetry, 
we need only consider an incoming right-traveling wave, 
for example, 
and all 
$\pm$ 
superscripts become unnecessary. 
Written in terms of phase shifts
$\d_\ell$, 
the asymptotic scattering solution 
becomes 
\begin{multline}
\psi(x) 
\overset{k|x| \gg 1}{=} 
e^{i \d_0} \, \cos \big( k |x| + \d_0\big) 
\\
+ i e^{i \d_1}  \, \sgn(x) \cos \big(k |x| - \tfrac{\p}{2} + \d_1\big)
\label{eq:pwa}
.\end{multline}
When both phase shifts vanish, 
we recover only the incident right-traveling wave, 
which justifies their interpretation as shifts in phase due to an interaction.  
Matching the two forms of the asymptotic solution in
Eqs.~\eqref{eq:f0f1} and \eqref{eq:pwa} 
leads to expressions for the partial-wave amplitudes in terms of the phase shifts
\begin{equation}
f_\ell 
= \frac{1}{\cot \d_\ell - i}
.\end{equation}
From the 
$T$-matrix, 
we obtain the 
$S$-matrix 
via 
Eq.~\eqref{eq:Tmatrix}, 
which is similarly diagonal on account of parity invariance
\begin{equation}
\mathbb{S} 
= 
\text{diag} 
\left( e^{2 i \delta_0}, e^{2 i \delta_1} \right)
\label{eq:Sell}
.\end{equation}
For the 
$S$-matrix to have unimodular eigenvalues, 
the 
$\d_\ell$
must be real valued.

Note that for a parity-even interaction, 
Eq.~\eqref{eq:RTfs} 
yields the relations 
\begin{equation}
f_0
= 
\frac{T + R - 1}{2i} 
\quad \text{and} \quad
f_1
=
\frac{T - (1 + R)}{2i}
\label{eq:TRP}
.\end{equation} 
In particular, 
the antisymmetric amplitude 
for a point interaction 
is proportional to the discontinuity of the wavefunction across the origin. 
From these amplitudes,  
we obtain the $S$-matrix elements in 
Eq.~\eqref{eq:Sell}
\begin{equation}
e^{2 i \d_0} = T + R
\quad \text{and} \quad 
e^{2 i \d_1} = T - R
\label{eq:phases}
,\end{equation}
which also directly follow from restricting
Eq.~\eqref{eq:SS}
to a parity-even interaction.
In Sec.~\ref{s:PE},
the machinery to diagonalize the 
$S$-matrix for a parity-even point interaction is unnecessary.
With 
$\a= \g$ 
and 
$\phi = 0$, 
the mathematical analogue of the magnetic field 
$\vec{\c B}$
in
Eq.~\eqref{eq:Bvec}
points in the third direction, 
which renders 
$\mathbb{S}$ 
diagonal. 
As a result of parity invariance, 
the amplitudes in 
Eq.~\eqref{eq:RT}
indeed satisfy
$R^{(+)} = R^{(-)} \equiv R$
and
$T^{(+)} = T^{(-)} \equiv T$, 
and the partial-wave phase shifts are identified as in Eq.~\eqref{eq:phases}. 
Using the expressions for 
$T$ 
and
$R$, 
moreover,
we recover the phase shifts in
Eq.~\eqref{eq:PEphases}.

\section{Regularization Schemes}
\label{s:DimReg}

The contact interactions in 
Sec.~\ref{s:C1D} requires regularization and renormalization. 
The scattering problem is formulated in momentum space, 
for which calculations require integrals of the form 
\begin{equation}
I_{2n}(k) = \int \frac{dp}{2\pi} \frac{p^{2n}}{p^2 - k^2 - i \epsilon}
\label{eq:I2n}
,\end{equation}
where 
$n \in \mathbb{Z}^+$
and 
$k > 0$
is the scattering momentum. 
The integral for 
$n = 0$
is convergent with
$I_0(k) = \frac{i}{2k}$, 
and must not be altered by the regularization scheme. 
For odd powers, 
the related integrals 
$I_{2n+1}(k)$
vanish in any parity-preserving regularization scheme. 
We employ two different regulators:
a hard momentum cutoff
and dimensional regularization. 
These are detailed below; 
and, 
for the latter, 
we employ two different schemes.

\subsection{Momentum Cutoff}

A straightforward way to tame ultraviolet divergences is to introduce a hard momentum cutoff
$\Lambda$, 
beyond which there are no momentum modes. 
This results in a simple definition of the regulated integrals in terms of the allowed modes
$|p| < \Lambda$, 
namely
\begin{multline}
I^\L_{2n}(k)
\equiv
\int_{-\L}^{+\L} \frac{dp}{2 \pi} \,
\frac{p^{2n}}{p^2 - k^2 - i \epsilon}
\\
=
k^{2n-1}
\Big[
b_0 
+ 
\sum_{j=1}^{n}
b_j \, \left(\tfrac{\L}{k}\right)^{2j-1} 
{+} \,
\c O \left(\tfrac{k}{\L}\right)
\Big]
\label{eq:ILambda}
,\end{multline}
where the maximum power of the cutoff is
$2n-1$, 
which is the degree of divergence of the integral. 
Note that logarithmic dependence on 
$\L$ 
will not be encountered. 
In the case of 
$n=0$, 
we have
$b_0(k) = \frac{i}{2}$
and the cutoff dependence of the integral is proportional to 
$\L^{-1}$. 
Taking 
$\L / k \gg 1$, 
we appropriately arrive at
$I_0^\L(k) = \frac{i}{2k}$.

To compute
$I_2^\L(k)$, 
note that the identity
\begin{equation}
\frac{p^2}{p^2 - k^2 - i \epsilon}
= 
\frac{k^2}{p^2 - k^2 - i \epsilon} + 1
\label{eq:reduce}
,\end{equation}
can be applied to the integrand. 
With the momentum cutoff regulating the integral, 
the integral of the sum is the sum of the integrals, 
namely
\begin{equation}
I_2^\L(k) = k^2 \, I^\L_0(k) +  I_2^{\L}(0) = \frac{ik}{2} + \frac{\L}{\p}
\label{eq:I2L}
.\end{equation}
The latter factor appearing above is the hard cutoff regularization of the delta function at the origin
\begin{equation}
\d(0) \longrightarrow I_2^\L(0) = \int_{-\L}^{+\L} \frac{dp}{2\p} \, 1 = \frac{\L}{\p}
\label{eq:d0L}
.\end{equation}
The identity 
Eq.~\eqref{eq:reduce} 
can be applied iteratively to reduce numerators containing higher powers of $p^2$. 
With a hard momentum cutoff, 
the derivative of the delta function at the origin is
\begin{equation}
\d'(0) \longrightarrow \int_{-\L}^{+\L} \frac{dp}{2 \p}\, i p = 0
,\end{equation}
and vanishes due to parity.

\subsection{Dimensional Regularization}

In dimensional regularization
(DR), 
divergent integrals are regulated by altering the number of dimensions, 
and then performing an analytic continuation in 
$d$. 
Logarithmic divergences show up as poles, 
which must be subtracted; 
whereas, 
power-law divergences are automatically renormalized. 
The power-divergence subtraction (PDS) scheme was devised to circumvent the automatic renormalization 
of power-law divergences%
~\cite{Kaplan:1998tg,Kaplan:1998we}. 
With parity-violating operators in one dimension there is an additional feature. 
The single derivative
$\frac{d}{dx}$ 
does not have an obvious generalization to 
$d$ dimensions. 
We employ two different schemes to handle this feature.

\subsubsection{Na\"ive Dimensional Regularization}

The first DR scheme is one that ignores features specific to one dimension. 
For this reason, 
we call it na\"ive DR (NDR). 
We define the dimensionally regulated integrals 
Eq.~\eqref{eq:I2n} 
in the na\"ive scheme as
\begin{equation}
I^\text{NDR}_{2n}(k) = \m^{1-d} \int \frac{d^d p}{(2\pi)^d}\frac{p^{2n}}{p^2 - k^2 - i \epsilon}
,\end{equation}
where we have introduced the 
DR scale 
$\m$
so that the integral in 
$d$ dimensions maintains the same physical units as in one dimension.
This definition merely replaces the one-dimensional momentum integral with a 
$d$-dimensional momentum integral. 
The integration ranges over the full $d$-dimensional momentum space.

Evaluation of the integral in $d$ dimensions yields
\begin{equation}
I_{2n}^\text{NDR}(k) 
= 
\frac{\p  i}{\sin \left( \frac{\p d}{2}  \right)} \,
\frac{\ k^{2n-1} \left( \frac{\m}{- i k\,} \right)^{1-d}}{(4 \p)^{d/2} \, \Gamma\big(\frac{d}{2}\big)}  
\label{eq:NDR}
,\end{equation}
having used that 
$n \in \mathbb{Z}^+$. 
Because there are no logarithmic divergences, 
the regulated integral is finite when evaluated in
$d = 1$. 
Thus, 
we have
\begin{equation}
I_{2n}^\text{NDR}(k) 
\overset{d=1}{=} 
\frac{i}{2} k^{2 n - 1}
\label{eq:INDR}
.\end{equation}
In particular, 
the result for 
$n = 0$
is 
$I^\text{NDR}_0(k) = \frac{i}{2k}$, 
which agrees with the direct evaluation of this finite integral.
For
$n = 1$, 
we have the result
$I^\text{NDR}_2(k) = \frac{ik}{2}$, 
which implies the regulated value for the delta function at the origin is
\begin{equation}
\delta(0) 
\longrightarrow  I_2^\text{NDR}(0) = 0
\label{eq:NDRd0}
.\end{equation}
This vanishing is similarly true of 
$\d'(0)$
(although due to parity),  
$\d''(0)$, 
\emph{etc}. 
The value of 
$I_2^\text{NDR}(k)$
can also be consistently found from the regulated recursion relation obtained from integrating 
Eq.~\eqref{eq:reduce}
\begin{equation}
I_2^\text{NDR}(k) = \delta(0) + k^2 I_0(k) = \frac{ik}{2}
\label{eq:I2NDR}
.\end{equation}
Compared to the hard momentum cutoff, 
the automatic subtraction of power-law divergences in 
NDR
leads to the relation
$I_{2n}^\text{NDR}(k) = I_{2n}^\L(k) \big|_{\L = 0}$.

\subsubsection{One-Dimensional PDS Scheme}

The NDR scheme is perfectly fine in most situations 
but potentially presents an issue for parity-odd operators, 
for example, 
those contributing to
$\c O_1$
in 
Eq.~\eqref{eq:O1}. 
These operators possess a classical scale symmetry, 
and the scale symmetry is respected within 
NDR. 
The regulated integrals in 
Eq.~\eqref{eq:INDR}, 
for example,
naturally maintain 
$\m$
independence. 
The quantum anomaly, 
however, 
should physically introduce a scale. 
The PDS scheme is ideal for introducing a renormalization scale by subtracting the divergence in one dimension lower;
however, 
the NDR integral in 
Eq.~\eqref{eq:NDR} 
does not have a pole in 
$d = 0$
dimensions.
Instead, 
it has the finite value
\begin{equation}
I_{2n}^\text{NDR}(k) \Big|_{d=0} 
= - \m \, (k^2)^{n-1} 
\label{eq:INDR0}
.\end{equation}

To apply a PDS scheme, 
we must be careful to extend the integrals to 
$d$ dimensions. 
Each factor of momentum in the numerator of the one-dimensional integral in
Eq.~\eqref{eq:I2n}
results from the action of 
$\frac{d}{dx}$. 
Extending the integrand to $d$ dimensions using NDR, 
factors of 
$p^2$
are treated as the magnitude-squared momentum.
To define our PDS scheme, 
the kinetic energy is taken as
$d$ dimensional, 
but the contact interactions are defined using only derivatives with respect to 
$x$, 
\emph{i.e}.~the first component of a $d$-dimensional position vector.  
As an illustration,
the Hamiltonian with only the derivative of the delta-function interaction in 
$d$
dimensions is taken to be
\begin{equation}
H= \frac{\ \vec{p} \, {}^2}{2m} - \frac{c_1}{m} \frac{d}{dx} \delta(\vec{r})
.\end{equation}
For this and similar interactions, 
we are thus led to define the integrals in DR as
\begin{equation}
I_{2n}^\text{DR}(k) = \m^{1-d} \int \frac{d^d p}{(2\pi)^d} \frac{(p_x)^{2n}}{p^2 - k^2 - i \epsilon}
.\end{equation}
Finally, 
the regulated value in the PDS scheme is obtained by subtracting the analytical form of the pole near 
$d =0$,
but evaluated in 
$d =1$
\begin{equation}
I^\text{PDS}_{2n}(k)
\equiv
\Big[ \ I_{2n}^\text{DR}(k) - I_{2n}^\text{DR}(k)\big|_{d\approx0} \, \Big]_{d=1}
\label{eq:IPDS}
.\end{equation}

To employ the 
PDS 
scheme to evaluate the 
DR 
integrals required in 
Sec.~\ref{s:C1D}, 
we first note that for 
$n =0$, 
the values are the same in DR and NDR,
$I^\text{DR}_0(k) = I^\text{NDR}_0(k)$. 
Because there is no pole in 
$I^\text{NDR}_0(k)$ 
as
$d$ 
nears zero, 
there is nothing to subtract in 
Eq.~\eqref{eq:IPDS}. 
Consequently, 
we obtain
$I^\text{PDS}_0(k) = \frac{i}{2k}$. 
Differences appear once 
$n \neq 0$.

For 
$n = 1$,  
appealing to 
$SO(d)$
rotational invariance allows us to relate the DR and NDR integrals
\begin{equation}
I^\text{DR}_{2}(k) = \frac{1}{d} \, I^\text{NDR}_2(k)
\label{eq:I2ROT}
,\end{equation}
and the former has a pole in 
$d = 0$ 
given the finite limit in 
Eq.~\eqref{eq:INDR0} of the latter. 
The value of the integral in the PDS scheme 
Eq.~\eqref{eq:IPDS}
is 
\begin{equation}
I_2^\text{PDS}(k)
=
\Big[ \, I_2^\text{DR}(k) + \frac{\m}{d} \, \Big]_{d=1} = \frac{ik}{2} + \m
\label{eq:I2PDS}
,\end{equation}
which should be compared with 
Eqs.~\eqref{eq:I2L} and \eqref{eq:I2NDR}. 
Note that in the PDS scheme, 
the simple identity in 
Eq.~\eqref{eq:reduce}
does not immediately appear in calculations. 
Only after appealing to 
$SO(d)$
rotational invariance can the identity be used in practice. 
This is being taken into account, 
however, 
by the relation in 
Eq.~\eqref{eq:I2ROT}.

\bibliographystyle{elsarticle-num-names} 
\bibliography{bibfile.bib}

\end{document}